\documentclass{jfm}
\usepackage{natbib}
\usepackage{hyperref}
\hypersetup{
    colorlinks = true,
    urlcolor   = blue,
    citecolor  = black,
}

\newcommand{\RomanNumeralCaps}[1]
\linenumbers
\usepackage{graphicx}
 \usepackage{changes}
\usepackage{epstopdf, epsfig}
\usepackage{newtxmath}
\usepackage{amsmath}
\usepackage{esdiff}
\usepackage{booktabs}
\usepackage{float}
\usepackage{arydshln}
\usepackage{placeins}
\DeclareMathOperator\erf{erf}

\shorttitle{Induced flow velocity from wake superposition}
\shortauthor{N. Mohebbi, J. Hwang, M. K. Fu and J. O. Dabiri}

\title{Measurements and modelling of induced flow in collective vertical migration}

\author{Nina Mohebbi\aff{1},
 Joonha Hwang \aff{1},
 Matthew K. Fu\aff{1},
 \and John O. Dabiri\aff{1,2} \corresp{\email{jodabiri@caltech.edu}}}

\affiliation{\aff{1}Graduate Aerospace Laboratories, California Institute of Technology,
Pasadena, CA 91125, USA
\aff{2}Mechanical and Civil Engineering, California Institute of Technology,
Pasadena, CA 91125, USA}

\begin{document}

\maketitle

\begin{abstract}
Hydrodynamic interactions between swimming or flying organisms can lead to complex flows on the scale of the group. These emergent fluid dynamics are often more complex than a linear superposition of individual organism flows, especially at intermediate Reynolds numbers. This paper presents an approach to estimate the flow induced by multiple swimmer wakes in proximity using a semianalytical model that conserves mass and momentum in the aggregation. The key equations are derived analytically, while the implementation and solution of these equations are carried out numerically. This model was informed by and compared with empirical measurements of induced vertical migrations of brine shrimp, \textit{Artemia salina}. The response of individual swimmers to ambient background flow and light intensity was evaluated. In addition, the time-resolved three-dimensional spatial configuration of the swimmers was measured using a recently developed laser scanning system. Numerical results using the model found that the induced flow at the front of the aggregation was insensitive to the presence of downstream swimmers, with the induced flow tending towards asymptotic beyond a threshold aggregation length. Closer swimmer spacing led to higher induced flow speeds, in some cases leading to model predictions of induced flow exceeding swimmer speeds required to maintain a stable spatial configuration. This result was reconciled by comparing two different models for the near-wake of each swimmer. The results demonstrate that aggregation-scale flows result from a complex, yet predictable interplay between individual organism wake structure and aggregation configuration and size. 
\end{abstract}


\clearpage

\section{Introduction}\label{Intro}

Various species of swimming and flying organisms exhibit collective motion, characterized by coordinated movement within groups of organisms \citep{Vicsek_Zafeiris_2012}. The emergent hydrodynamic properties of collective groups of swimming and flying organisms are vital to understanding flow-mediated communication \citep{Mathijssen_Culver_Bhamla_Prakash_2019}, fluid transport \citep{Katija_2012}, and the hydrodynamic performance of collectives \citep{Weihs_1973,Zhang_Lauder_2023}. Applications of these fluid mechanics include control mechanisms for robotic swarms \citep{Berlinger_Gauci_Nagpal_2021} and climate modelling \citep{Stemmann_Boss_2012}. 

One of the most common manifestations of collective behaviour found in the ocean is diel vertical migration (DVM). Prevalent among freshwater and marine zooplankton taxa globally, DVM involves the migration of zooplankton from deep regions in the water column during the day to shallower depths at night over a vertical distance of the order of 1 km; it is the largest migration on Earth by mass \citep{Bandara_Varpe_Wijewardene_Tverberg_Eiane_2021}. However, the scale of flow induced by a DVM event remains unresolved despite numerous field measurements \citep{Fernández_Castro_Peña_Nogueira_Gilcoto_Broullón_Comesaña_Bouffard_Naveira_Garabato_Mouriño-Carballido_2022, Dewar_Bingham_Iverson_Nowacek_Laurent_Wiebe_2006, Farmer_Crawford_Osborn_1987, Gregg_Horne_2009}, laboratory observations \citep{Houghton_Koseff_Monismith_Dabiri_2018}, and theoretical estimates \citep{Huntley_Zhou_2004, Dewar_Bingham_Iverson_Nowacek_Laurent_Wiebe_2006} of biogenic mixing due to collective swimming.

Studies of flows on the individual organism scale include a comprehensive set of experimental \citep{Lauder_Madden_2008, Dabiri_2005}, theoretical \citep{Derr_Dombrowski_Rycroft_Klotsa_2022, Wu_2011} and computational \citep{Pedley_Hill_1999, Eldredge_2007} estimates. Direct numerical simulation has been used to study the hydrodynamics of collective motion \citep{Ko_Lauder_Nagpal_2023}, including the mixing induced by Stokes squirmers \citep{Ouillon_Houghton_Dabiri_Meiburg_2020,Wang_Ardekani_2015,Lin_Thiffeault_Childress_2011}. However, due to the nonlinear coupling between individual and collective flow fields at intermediate and high Reynolds numbers, connecting these individual flows to the fluid dynamics on the collective scale remains an open challenge using a modelling approach short of direct numerical simulation. 

At low Reynolds numbers, Stokesian dynamics \citep{Brady_Bossis_1988} can be used to estimate hydrodynamic interactions through linear superposition \citep{Ishikawa_Simmonds_Pedley_2006,Pushkin_Shum_Yeomans_2013,Lauga_Powers_2009}. For organisms characterized by high Reynolds number dynamics, the linearity of potential flow theory allows for approaches based on linear superposition to estimate the combined effect of flow within a group \citep{Weihs_1973,Weihs_2004}. However, for swimmers operating in an intermediate Reynolds regime, such as the majority of vertically migrating swimmers in the ocean \citep{Katija_2012}, neither Stokesian nor potential flow assumptions accurately capture the dominant hydrodynamic forces, resulting in nonlinear governing dynamical equations that are not readily suitable for linear superposition.

Although not strictly justified from first principles, superposition has been successfully applied to estimate wake interactions in wind farms without using potential flow assumptions. Initial efforts, exemplified by the linear superposition model proposed by \citet{Lissaman_1979}, assumed a large wind turbine spacing and weak wake interactions to linearly sum wake velocity deficits. Subsequent critiques highlighted the potential overestimates of the wake deficit within densely arranged wind turbine arrays, where there are significant wake interactions \citep{Crespo_Hernández_Frandsen_1999}. In response to this limitation, several alternative superposition methods have been proposed. \citet{Katic_Højstrup_Jensen} posited that the combined velocity deficit in the wake overlap regions can be estimated by a sum of the squares of individual velocity deficits. \citet{Voutsinas_Rados_Zervos_1990} proposed a model that assumes that the total energy loss in the superposed wake is equal to the sum of the energy losses of each turbine upwind. Each of the aforementioned models demonstrated improved agreement with the measurement data, especially with stronger wake interactions. However, each model lacks a theoretical justification based on the conservation of mass and momentum in the wake. Recently, \citet{Zong_Porté-Agel_2020} introduced a model that explicitly conserves mass and momentum in regions of wake overlap. This approach demonstrated superior performance over previous models compared with experimental and large-eddy simulation data.

Here, we adapt the approach of \citet{Zong_Porté-Agel_2020} to develop an analytical model that estimates the three-dimensional (3-D) flow induced by wake interactions of swimmers using brine shrimp as a model organism.  The model was developed to conserve mass and momentum, drawing empirical parameters from the swimming trajectories of brine shrimp during induced vertical migration. We introduced an estimated convection velocity term to calculate mass flux in a linearized momentum equation. This was used to develop an analytical wake superposition model based on each swimmer's local flow and the geometric configuration of the collective group (\S \ref{sec:Analytical}). The swimming trajectories of brine shrimp were measured (\S \ref{sec:individuals}, \ref{sec:individuals_2}, \ref{sec:collective}) to discern the effects of environmental variables on the behaviour of individual swimmers (\S \ref{sec:results_individual}, \ref{sec:results_collective}). These empirical findings informed the parameters used in the computational model (\S \ref{sec:model_experiment}). We found that the aggregate-scale induced flow was a function of the individual wake shape, length of the group, and animal number density. In addition, we found that the induced flow can be significantly stronger than the flow associated with individual swimmers (\S \ref{sec:results_model}).

\section{Analytical model}\label{sec:Analytical}

\subsection{Individual swimmer wake model}\label{sec:Linearized}

This section introduces an analytical model to compute the flow field generated by many individual wakes in close proximity while conserving mass and momentum. This method is inspired by the approach adopted by \citet{Zong_Porté-Agel_2020} to superpose wind turbine wakes. Unlike previous formulations, which prescribe a drag coefficient and calculate momentum deficits, the present formulation prescribes the net force generated by the swimmers and calculates momentum excess. Importantly, this formulation did not assume a priori that the convective velocity would trend towards a plateau.

We assume that a vertical swimmer generates a downstream wake defined in the swimmer-fixed frame $u_w(x,y,z)$ to generate a net force $F_z$ that counteracts negative buoyancy and thus maintains a constant swimming speed $u_0$ through a fluid with constant density $\rho$. These assumptions allow for the simplification of the integral form of the momentum equation,

\begin{equation}
F_z = \rho \iint_{wake} {u_w(x,y,z)} (u_w(x,y,z)-u_0) \;dxdy .\label{eqn:F_uw_u0}
\end{equation}

By introducing the wake velocity surplus, $u_s = u_w-u_0$, and substituting this definition into equation \ref{eqn:F_uw_u0} we obtain the following:

\begin{equation}
F_z = \rho \iint_{wake} {u_w(x,y,z)} u_s(x,y,z) \;dxdy. \label{eqn:basic_momn}
\end{equation}

We introduce an effective wake convection velocity, $u_c(z)$, which varies with downstream distance from the swimmer, but is constant in the spanwise directions. Consequently, the net vertical force can be rewritten as

\begin{equation}
F_z = \rho {u_c(z)} \iint_{wake} u_s(x,y,z) \;dxdy. \label{eqn:F_uc}
\end{equation}

The wake convection velocity effectively represents the average speed at which the local velocity surplus is advected in the wake of the swimmer. To derive a mathematical expression for $u_c$, we substitute \ref{eqn:F_uc} into \ref{eqn:basic_momn} to get

\begin{equation}
{u_c(z)} = \frac{\displaystyle\iint_{wake} {u_w(x,y,z)} u_s(x,y,z) \;dxdy}{\displaystyle\iint_{wake} u_s(x,y,z) \;dxdy}. \label{eqn:uc}
\end{equation}

The numerical evaluation of equation \ref{eqn:uc} is described in \S\ref{wake_equations}.

\subsection{Wake superposition} \label{sec:superposition}

To calculate the flow field at the aggregate scale, we define $U_{\infty}$ as the swimming speed of all organisms in the volume, or the free stream velocity in a swimmer-fixed frame. Furthermore, we introduce $U_w(x,y,z)$ as the global flow field generated by the swimmers. Lastly, $U_s$ is defined as the velocity surplus generated by the swimmers expressed as $U_s(x,y,z) = U_w(x,y,z) - U_{\infty}$. Following a procedure analogous to that in \S\ref{sec:Linearized}, the effective convection velocity of the combined wakes is given by the following: 

\begin{equation}
{U_c(z)} = \frac{\displaystyle\iint {U_w(x,y,z)} U_s(x,y,z) \;dxdy}{\displaystyle\iint U_s(x,y,z) \;dxdy}. 
\label{eqn:Uc}
\end{equation}

The force exerted by the $i$th swimmer in the streamwise direction is denoted $F_z^i$. To conserve momentum in the wake, we require

\begin{equation}
\sum_{i} F_z^i = \rho {U_c(z)} \iint U_s(x,y,z) \;dxdy. \label{eqn:sumF_Uc}
\end{equation}

Substituting the left-hand side of \ref{eqn:sumF_Uc} with \ref{eqn:F_uc} yields the following:

\begin{equation}
\sum_{i} \rho {u_c^i(z)} \iint u_s^i(x,y,z) \;dxdy = \rho U_c(z) \iint U_s(x,y,z) \;dxdy. \label{eqn:sum_momn}  
\end{equation}

The velocity experienced by the $i$th organism is denoted $u_0^i$ and defined as $U_w(x^i,y^i,z^i)$ based on upstream swimmers. The wake velocity induced by the $i$th organism is $u_w^i$, and the wake velocity surplus for the $i$th organism, $u_s^i$, is expressed as $u_w^i-u_0^i$. The rearrangement of these terms and the subsequent application of the analysis across the entire volume lead to the derivation of an expression for the global wake surplus,

\begin{equation}
U_s(x,y,z) = \sum_{i}\frac{u_c^i(z) }{U_c(z)} u_s^i(x,y,z).
\label{eqn:Us}
\end{equation}

In light of \ref{eqn:Uc}, which describes $U_c$ as a function of $U_s$ and \ref{eqn:Us}, which characterizes $U_s$ as a function of $U_c$, an iterative methodology is used to solve for $U_s$ and $U_c$. The procedure begins with the assumption $U_c=U_\infty$, where $U_\infty$ denotes the velocity of the free stream. This is an underestimate, as $U_c$ will increase from the free stream velocity with added momentum provided from the swimmers. Thus, in the first iteration, \ref{eqn:Us} is used to evaluate $U_s$, and will result in an overestimate since its value is inversely related to that of $U_c$. As the iterative process continues, this overestimate of $U_s$ is used in \ref{eqn:Uc} to refine the calculation of $U_c$, increasing the estimate from the initial guess. In this way $U_c$ will continue increasing from the initial guess and $U_s$ will continue decreasing until the value of $U_c$ converges, satisfying condition $|U_c-U_c^*|/U_c^* \leq \epsilon$, where $U_c$ is calculated from the preceding iteration, $U_c^*$ is calculated from the ongoing iteration, and $\epsilon=0.01$. The iterative development of local and global estimated convective velocities captures inherent nonlinearity and ensures the conservation of momentum in the establishment of the final 3-D flow field.

\section{Experimental methods} \label{sec:experimental_methods}
Experiments with brine shrimp, \textit{Artemia salina}, which swim at a Reynolds number around 100, provided a model for planktonic vertical migrations at intermediate Reynolds numbers. As demonstrated in previous work \citep{Houghton_Koseff_Monismith_Dabiri_2018,Fu_Houghton_Dabiri_2021}, brine shrimp exhibit a phototactic response, swimming towards a nearby light source. This facilitates controllable vertical migrations in a laboratory setting. The flow and light intensity encountered by an individual swimmer depend on its specific location within the collective. Therefore, the dynamics of each swimmer in aggregation were anticipated to depend on the local light stimulus and ambient flow. Consequently, \S\ref{sec:individuals} describes experiments designed to characterize the response of brine shrimp to varying light stimuli and background flows. In \S\ref{sec:collective}, we detail the techniques developed to measure 3-D reconstructions of swimming trajectories, aiming to establish the relationship between the number of swimmers migrating and the average nearest neighbour distance, a descriptor of the swimmer configuration. Finally, \S\ref{sec:model_experiment} uses the insights gained from these experiments to set the modelling parameters and formulate numerical simulations of the flow induced by collective vertical migration. 

\subsection{Individual swimmer response to light stimulus} \label{sec:individuals}

\begin{figure}
 \centering{\includegraphics[scale=0.29]{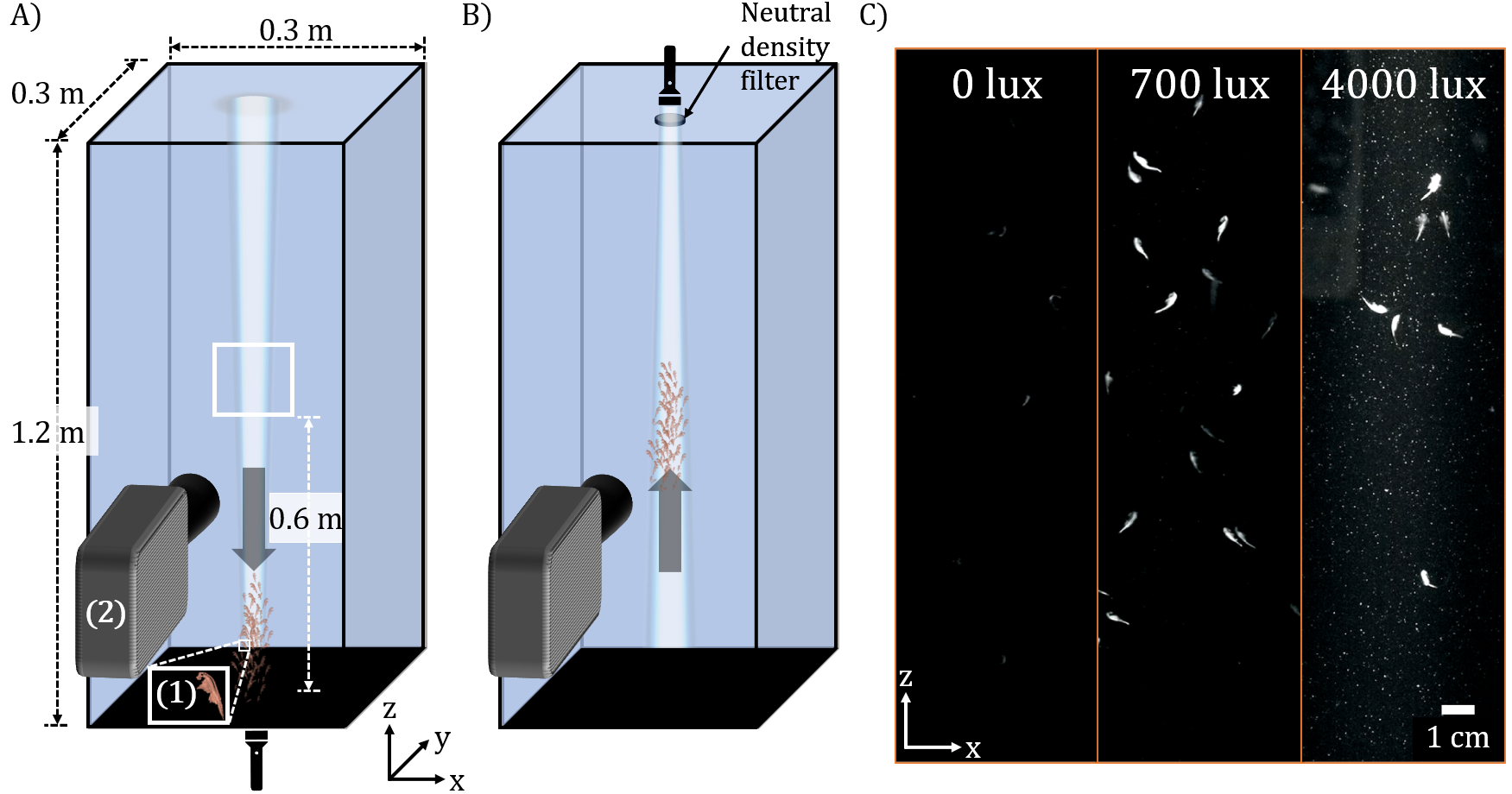}}
 \caption{Schematic of experimental protocol for characterization of phototactic response of brine shrimp. A) Brine shrimp (1) were gathered at the bottom of a 1.2 m tall tank using a flashlight positioned at the base. B) To induce vertical migration, the bottom flashlight was turned off, and the top flashlight was turned on. The light intensity of the top flashlight was varied using neutral density filters. Recording was manually initiated once the swimmers entered the field of view of the high-speed camera (2). C) Example frames from the x-z plane captured during vertical migrations under different light intensities, adjusted with neutral density filters. All trials were conducted with an infrared tank illuminator, which was used exclusively in the 0 lux condition. Three trials were performed for each light intensity: 0, 800, 1500, 2300, and 4000 lux. }
\label{fig:exp_light}
\end{figure}

The response of brine shrimp to different light intensities was investigated in a controlled environment. A 1.2 m high tank with a cross section of 0.3 m x 0.3 m (Figure \ref{fig:exp_light}) was filled with 35 parts per thousand of salt water using Instant Ocean Sea Salt (Spectrum Brands). To reduce the influence of swimmer wakes on one another, the tank was populated with less than 1500 swimmers, or 0.015 animals per cm$^3$. All experiments were carried out within 24 hours of animal acquisition.

To ensure consistency between trials, the animals were gathered at the bottom of the tank using a flashlight, and a minimum settling time of 30 minutes was allowed between each trial. Initiating a vertical migration involved turning off the flashlight at the tank's bottom and activating a target flashlight (PeakPlus LFX1000, 1000 lumens) positioned above the tank. The light intensity of this upper flashlight was adjusted using three different neutral density filters: 1/2, 1/4, and 1/8 transmittance (Neewer 52mm ND Filter Kit). A light intensity meter (TEKCOPLUS Lux Meter with Data Logging) was used to measure the illumination at the bottom of the tank for each filter.

A high-speed camera (Edgertronic SC1) was set up with a 20 cm x 25 cm (1024 pixel x 1280 pixel) field of view, 60 cm above the bottom of the tank. For each test, a recording was manually triggered once the first swimmer entered the camera field of view and captured for 30 seconds at 40 frames per second (fps). Four trials were carried out with each of the three filters (800, 1500, 2300 lumens per square meter(lux)), without a filter present (4000 lux), and without the target light (0 lux). An infrared (850 nm) light was used to illuminate the tank and collect control data for the case in which no visible illumination was present. For consistency across all tests, this infrared illumination remained on for all tests. 

\subsection{Individual swimmer response to background flow} \label{sec:individuals_2}
\begin{figure}
 \centering{\includegraphics[scale=0.29]{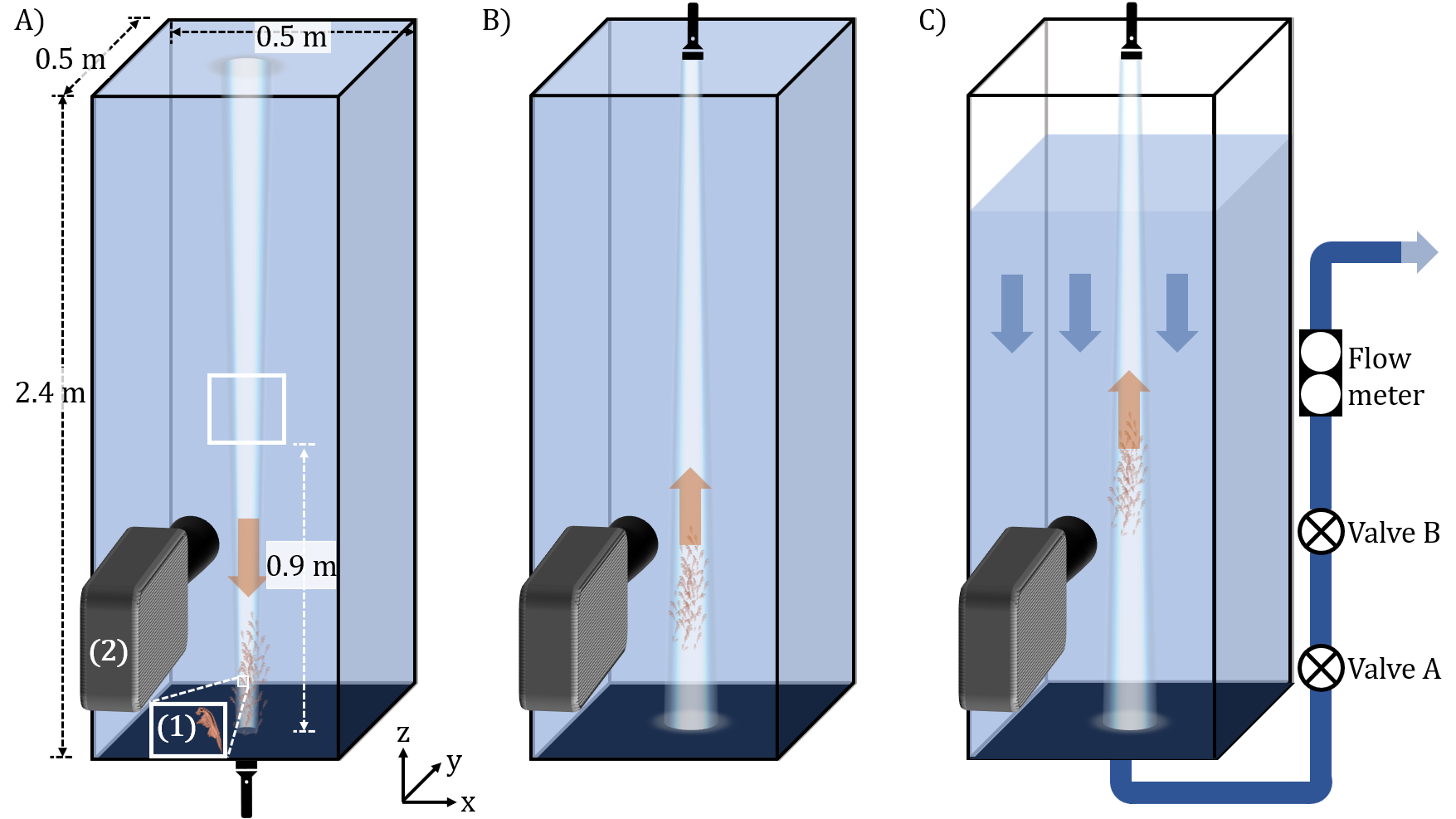}}
 \caption{Schematic of the experimental protocol for characterizing the flow response of brine shrimp. A) Brine shrimp (1) were initially gathered at the bottom of a 1.2 m tall tank using a flashlight positioned at the base. B) Brine shrimp (1) were initially gathered at the bottom of a 1.2 m tall tank using a flashlight positioned at the base. C) Once the swimmers entered the field of view of the high-speed camera (2), a flow valve was opened, introducing bulk flow in the opposite direction of the swimmers' motion. Recording was manually initiated at this point. The flow rate was controlled using a system of two flow valves and a flow meter arranged in series. Three trials were conducted for each target flow speed: 0, 0.07, 0.14, 0.21, and 0.3 cm s$^{-1}$.}
\label{fig:exp_flow}
\end{figure}

To simulate the vertical flows induced during collective vertical migration of brine shrimp, water was drained from a 2.4 m tall tank with a cross section of 0.5 m x 0.5 m, producing uniform flows in the range of 0.05-0.5 cm s$^{-1}$ (figure \ref{fig:exp_flow}). These flow speeds correspond to those observed in vertical migrations of brine shrimp, with animal number densities between 100,000 and 600,000 animals per cubic meter \citep{Houghton_Dabiri_2019}. The uniformity produced by this set-up reflects the uniform jet produced by steadily moving dilute swarms in discrete swimmer simulations \citep{Ouillon_Houghton_Dabiri_Meiburg_2020}.

Before testing, the tank was filled with 10 $\mu$m silver-coated glass spheres (CONDUCT-O-FIL, Potters Industries, Inc.) to facilitate imaging of the flow field with a laser sheet. To confirm the quiescence of the tank, particle image velocimetry (PIV) was employed after introducing the animals with a 15 mL centrifuge tube. The tank was considered quiescent when the maximum time-averaged streamwise velocity was below 0.02 cm s$^{-1}$. Flow rate control was achieved using two series connected flow valves (1 in. NPT PVC Ball Valve), one for flow control and one for shut-off, and an inline flow meter (FLOMEC Flowmeter/Totalizer 5-50 gpm). 

Once the tank was confirmed to be quiescent with PIV, a migration was induced with the same procedure explained in \S\ref{sec:individuals}. A high-speed camera (Edgertronic SC1) was set up with a field of view of 21 cm x 26 cm (1024 pixel x 1280 pixel), 90 cm up from the bottom of the tank. Once the first swimmer entered the camera field of view, the shut-off valve was manually opened to initiate the flow, and the camera was manually triggered to record for 30 seconds at 15 fps. Three trials were carried out for each of the five target speeds: 0, 0.07, 0.14, 0.21 and 0.3 cm s$^{-1}$. The trials were carried out on different days using different animals, considering the limited number of trials achievable with the volume of the tank. 

To identify potential sources of measurement uncertainty, two significant factors were addressed. First, before the onset of the flow, a streamwise velocity variation of 0.02 cm s$^{-1}$ was allowed. Second, the flow speed was manually set using a ball valve and changed during each test due to variations in the height of the water column during draining. To address these uncertainties, an additional camera recorded flow rates displayed on the inline flow meter during each test. Both of these sources of variability were accounted for in the error bars of all flow measurements.

\begin{figure}
 \centering{\includegraphics[scale=0.29]{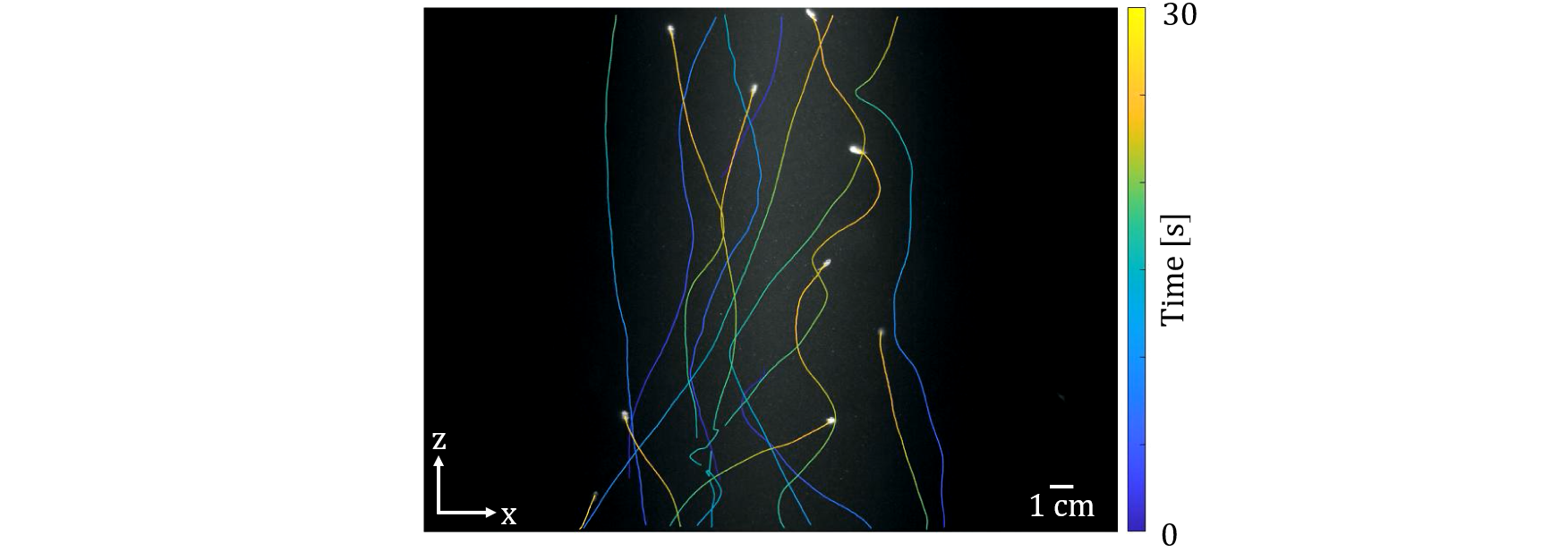}}
 \caption{Plot of brine shrimp swimming trajectories generated with ImageJ's wrMTrck plugin over a 30-second interval on top of the final frame in the image sequence. Colour transitions from blue to yellow represent progression in time, illustrating the trajectory of each swimmer within the tank. Brine shrimp present in the final frame can be identified as the white silhouettes at the end of the trajectories. Gradations in the background shading are due to the illumination used to induce phototaxis.}
\label{fig:2D_tracks}
\end{figure}

The captured videos of the vertical migration of brine shrimp were analysed using FIJI \citep{Schindelin_Arganda-Fiji_2012} and the wrMTrck plugin \citep{Husson_wrmtrck_2012}. The resulting swimming trajectories were fitted in MATLAB with a smoothing spline algorithm, which minimizes a combination of squared residuals and curvature penalties utilizing cubic smoothing spline interpolation to fit a curve to the provided data points. A smoothing parameter of 0.95 (figure \ref{fig:2D_tracks}) was selected to prioritize the reduction of oscillations, effectively smoothing out the fitted curve while preserving the overall trend of the data.

\subsection{Characterizing collective swimming} \label{sec:collective}

\begin{figure}
 \centering{\includegraphics[scale=0.29]{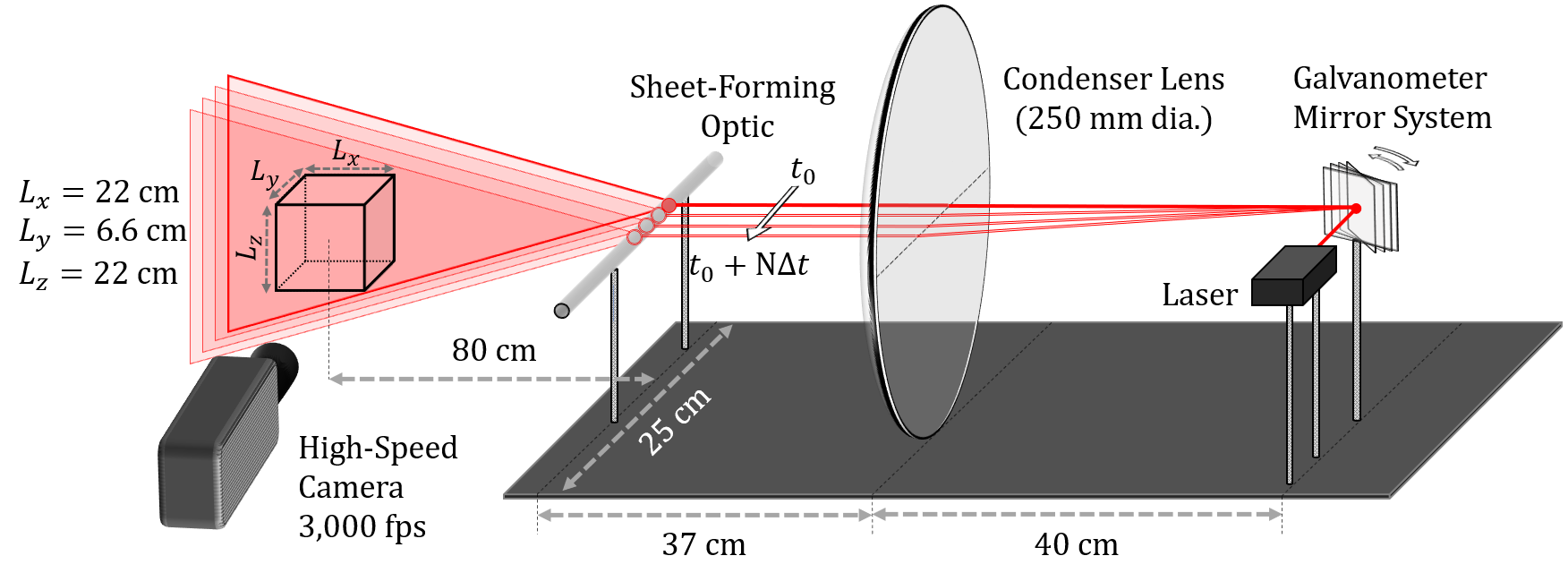}}
 \caption{Schematic of the scanning system, modified from \citet{Fu_Houghton_Dabiri_2021}. The laser beam was directed at a mirror whose angle was controlled by a galvanometer. Following reflection by the mirror, the beam was passed through a condenser lens and glass rod in order to generate a laser sheet parallel to the camera field of view.}
\label{fig:3DPTV}
\end{figure}

To reconstruct the 3-D swimming trajectories of brine shrimp during induced vertical migration, a 3-D particle tracking velocimetry (PTV) method was used (figure \ref{fig:3DPTV}), using scanning optics and a single high-speed camera (Photron FASTCAM SA-Z). A 671 nm continuous wave laser (5 W Laserglow LRS0671 DPSS Laser System) was directed through a condenser lens (370 mm back focal length) and a sheet-forming glass rod by a mirror to ensure parallel beams. The distance between the laser plane and the high-speed camera was adjusted with a galvanometer (Thorlabs GVS211/M) controlled by a voltage signal from an arbitrary function generator (Tektronix AFG3011C). This experiment was validated on a smaller scanning volume in the same facility with the same equipment by \citet{Fu_Houghton_Dabiri_2021}.

Each laser sheet sweep covered 6.6 cm of tank depth and took 0.1 seconds to complete. The high-speed camera captured 300 two-dimensional (2-D) 22 cm x 22 cm (1024 pixel x 1024 pixel) slices during this period. The scanned volume was centred on the tank cross-section and positioned 0.9 m from the tank floor. The swimmers move at approximately 1 cm s$^{-1}$; therefore, during the length of a scan (0.1 seconds), the swimmers will have moved approximately 0.1 cm. Given their body size of 1 cm, we effectively treated each scan as a still frame for the purposes of the current analysis. In addition, this combination of scanning rate and camera frames per second results in approximately 48 image sheets per cm in the scanning direction ($y$) and 46 pixels per cm in the camera plane ($x$ and $z$). This level of resolution for detecting swimmers 1 cm long results in well-formed and easily identifiable swimmers, as shown in Figure \ref{fig:3D_tracks}A. A video of 3-D reconstructed swimmers for 6 seconds is in the supplementary materials.

The experiments were carried out in the tank described in \S\ref{sec:individuals}. The brine shrimp were added to the tank in densely packed 0.25 teaspoon increments (approximately 125 swimmers) and three vertical migrations were induced as explained in \S\ref{sec:individuals} for each increment of swimmers added. Throughout vertical migration, the centre of the tank was scanned for 6 s every 50 s, totalling 7.5 minutes, to assess configuration changes over time.

A 3-D volume was constructed from 295 2D slices for each laser sweep (figure \ref{fig:3D_tracks}A). A custom MATLAB script was used to segment the 3-D volume and identify centroids. The volumetric data was median and Gaussian filtered to reduce noise and enhance object visibility. The filtered data was then binarized and morphological operations were applied, including opening with a spherical structuring element and hole filling, to refine the binary mask. This preprocessing ensured a clean and noise-reduced dataset for object detection. Through object detection, properties such as centroid, principal axis, and volume were identified for each connected component within the volume. Subsequently, a forward and backward nearest neighbour search in time was applied to centroid locations to identify and label swimming trajectories (figure \ref{fig:3D_tracks}B).

\begin{figure}
 \centering{\includegraphics[scale=0.29]{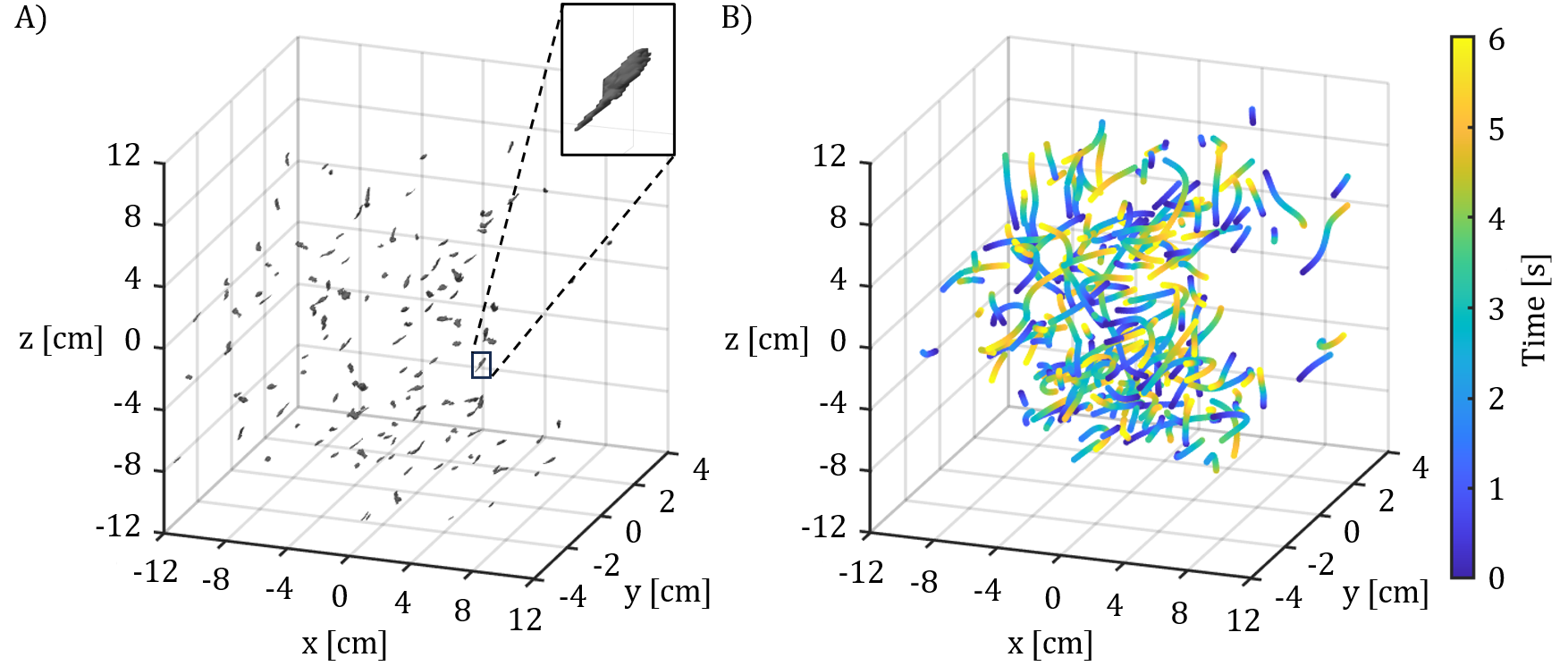}}
 \caption{Volume reconstruction and animal tracking during an induced vertical migration in the positive z direction. A) 3-D scanned reconstruction of 100 animal bodies. A close-up of an individual animal shown for reference. B) Six-second 3-D swimming trajectories, with colour transition from blue to yellow represent progression in time, illustrating the trajectory of each swimmer within the tank.}
\label{fig:3D_tracks}
\end{figure}

\subsection{Modelling assimilation} \label{sec:model_experiment}

\subsubsection{Wake profile models} \label{wake_equations}
Two models for the individual wake structure were studied to explore the impact of local flow geometry on the aggregation-scale flow. The local flow was defined as a function of the radial distance, $r = \sqrt{x^2 + y^2}$, and the characteristic width of the wake, $\sigma (z)$, at each value of z. First, a Gaussian model was implemented, consistent with the wake models previously used for wind turbine modelling,
\begin{equation}
  \xi_{gaussian}\left(\frac{r}{\sigma(z)}\right)= e^{-\frac{r^2}{2\sigma(z)^2}}. \label{xi_gaus}
\end{equation}
Second, a Ricker wavelet model was used to represent a local flow both in the direction of swimming and in the opposite direction of swimming,

\begin{equation}
  \xi_{wavelet}\left(\frac{r}{\sigma(z)}\right)= -\left(1 - \frac{r^2}{2\sigma(z)^2}\right) \: e^{-\frac{r^2}{3\sigma(z)^2}}. \label{xi_wave}
\end{equation}

\begin{figure}
 \centering{\includegraphics[scale=0.29]{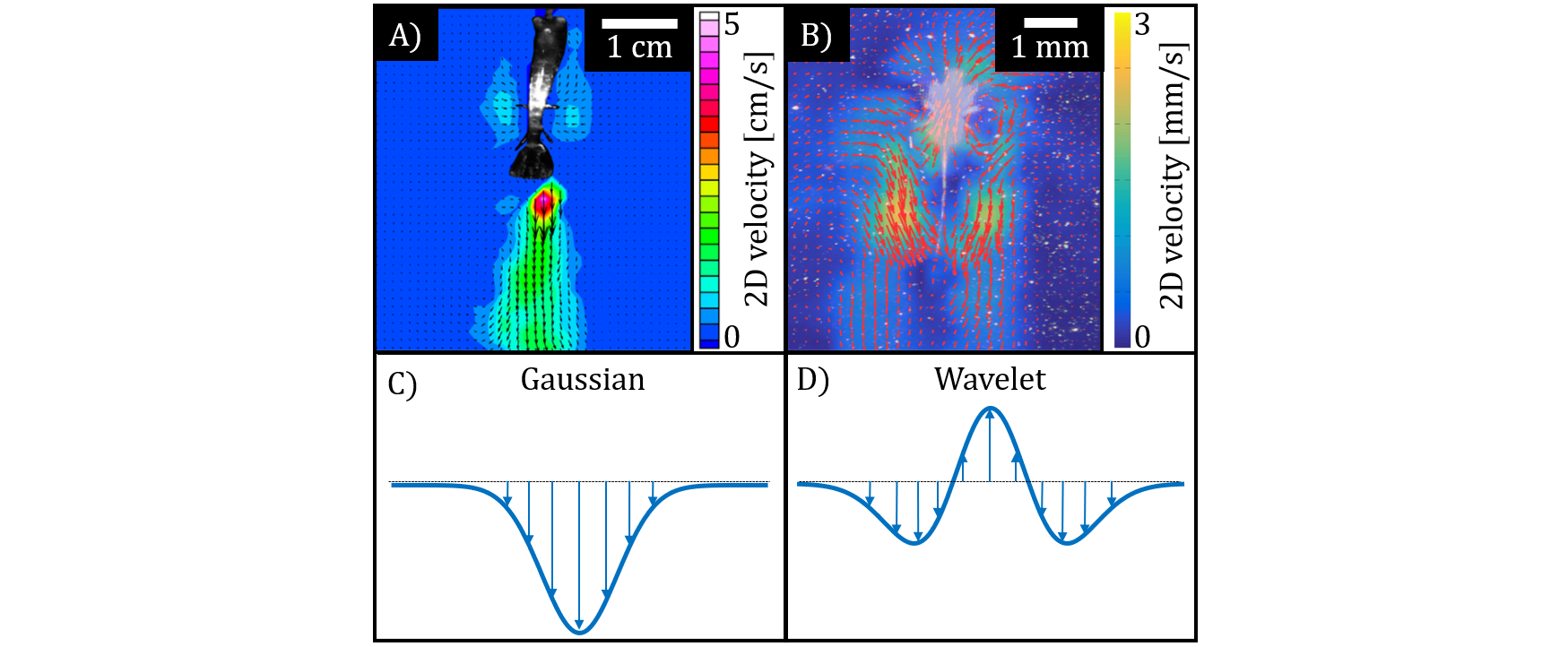}}
 \caption{Comparative analysis of flow fields generated by different swimmer types and corresponding wake models. A) The PIV results showing the flow field generated by a single free-swimming Pacific krill, adapted from \citet{Catton_Webster_Kawaguchi_Yen_2011} with the permission of the \textit{Journal of Experimental Biology}. The colour map represents the velocity magnitude, with arrows indicating flow direction. B) PIV results for a brine shrimp, highlighting the flow characteristics generated by its swimming motion, adapted from \citet{Wilhelmus_Dabiri_2014}, with the permission of AIP Publishing. C) A schematic representation of how a Gaussian distribution qualitatively captures the flow behind Pacific krill, which utilizes a single primary propulsor. (D) A schematic representation of a wavelet model that qualitatively captures the flow behind a brine shrimp, characterized by two sets of propulsors and a drag region immediately behind the body that induces backflow. The panels illustrate the differences in flow structures arising from distinct swimming mechanisms and motivate the comparison between a Gaussian and wavelet wake superposition model.}
\label{fig:methods_individual_PIV}
\end{figure}

These models will be referred to as the Gaussian and wavelet models, respectively. The Gaussian model, commonly employed in wind turbine modelling \citep{Zong_Porté-Agel_2020}, represents the flow behind the swimmer as a single downward jet. As evidenced by the qualitative match between figure \ref{fig:methods_individual_PIV}A  and the schematic in figure \ref{fig:methods_individual_PIV}C, the Gaussian wake function effectively captures the flow characteristics behind a swimmer utilizing a single main propulsor, generating a distinct single-lobed jet. The wavelet model, derived from the modified Ricker wavelet, is based on the second derivative of a Gaussian function, with an adjusted prefactor in the exponent denominator in order to create a function with a non-zero integral. As illustrated by the quantitative similarity between \ref{fig:methods_individual_PIV}B and the schematic in figure \ref{fig:methods_individual_PIV}D, the wavelet wake function captures the flow distribution generated by a swimmer with two sets of propulsive appendages, producing a double-lobed jet and a region of backflow immediately behind the swimmer due to the drag created by the body.

We modelled the spatial evolution of the propulsive jet as a self-similar, axisymmetric jet,

\begin{equation}
  u_s = u_0 \;c(z)\;\xi\left(\frac{r}{\sigma(z)}\right), \label{eqn:u_s}
\end{equation}

where $c(z)$ is a scaling factor and the shape of the wake is determined by $\xi(r/\sigma(z))$. An expression for $c(z)$ that conserves momentum by definition was derived by prescribing $u_0$, $\xi(r/\sigma(z))$, and $F_z$ (detailed steps provided in the Appendix \ref{appA}): 

\begin{equation}
  u_{s,gaussian}(x,y,z)= u_0 \left(1 - \sqrt{1 + \frac{F_{z}}{\pi \rho \sigma(z)^2 u_0^2}} \right) e^{-\frac{r^2}{2\sigma(z)^2}}. \label{final u_s gaussian}
\end{equation}

\begin{equation}
  u_{s,wavelet}(x,y,z)= u_0 \left( \frac{4}{5}-\frac{4}{5}\sqrt{1 + \frac{5 F_{z}}{12 \pi \rho \sigma(z)^2 u_0^2}}\right) \left(-1 \left(- \frac{r^2}{2\sigma(z)^2}\right) \: e^{-\frac{r^2}{3\sigma(z)^2}}\right), \label{final u_s wavelet}
\end{equation}

Substituting \ref{final u_s wavelet} and \ref{final u_s gaussian} into \ref{eqn:uc} and integrating in the streamwise direction over a circular cross section with infinite radius, the expressions for $u_c$ is obtained

\begin{equation}
u_{c, gaussian}(z) = \frac{u_0}{2} \left(1+ \sqrt{1+ \frac{F_z}{\pi \rho \sigma(z)^2 u_0^2}}\right). \label{final u_c gaussian}
\end{equation}

\begin{equation}
u_{c, wavelet}(z) = \frac{u_0}{2} \left(1+ \sqrt{1+ \frac{5 F_z}{12 \pi \rho \sigma(z)^2 u_0^2}}\right), \label{final u_c wavelet}
\end{equation}

A swimmer moving vertically at constant velocity must overcome the negative buoyancy that arises from the swimmer having a greater density, $\rho_s$, than sea water. The balance of force on the swimmer is expressed as follows:

\begin{equation} 
  F_{thrust}= g V_s (\rho_s-\rho)
\end{equation}

The thrust is introduced over some distance and not at an exact point in the flow. Therefore, we amended this expression to

\begin{equation} 
  F_{thrust}(z)= g V_s (\rho_s-\rho) \frac{1 + \erf \left(\frac{z}{L_s}\right)}{2}, \label{F_ext}
\end{equation}

for a more gradual development of the wake, where $L_s$ is the length of the swimmer's body length (BL). Similarly, an empirical model for the effective diameter of the wake as a function of the streamwise distance from the swimmer was used to capture the wake expansion,

\begin{equation} 
  \sigma (z)= 0.25 + 0.25 \ \text{log}(1+e^{(z-1.5)/L_s}) 
\end{equation}

The variables explicitly defined for the numerical implementation of this wake model are listed in table \ref{tab:fit_values}. These values were approximated to be of the order of observations made during laboratory experiments using brine shrimp. We normalize all length measurements by the BL of a swimmer, $L_s$. 

\begin{table}
  \centering
  \begin{tabular}{cccc}
     \toprule
      Symbol & Variable & Value & Unit\\
    \midrule
      $U_{\infty}$  & swimming velocity & 1 & {cm}/{s} \\
      $g$  & gravitational acceleration  & 9.8 & {m}/{s$^2$} \\
      $V_s$  & swimmer volume & $0.2$  & cm$^3$ \\
      $\rho_s$  & swimmer density & 1055 & {kg}/m$^3$ \\
      $\rho$  & seawater density  & 1025 & kg/m$^3$ \\
      $L_s$& body length& 1 & cm \\
    \bottomrule
  \end{tabular}
  \caption{Variables used in wake superposition model for induced flow in vertical migration.}
  \label{tab:fit_values}
\end{table}

\subsubsection{Collective flow field calculations}

Model flow fields for various swimmer configurations were calculated to identify the impact of aggregate characteristics on induced flow. First, changes due to group length were examined. The length of the group was increased in each test, while animal number density and width remained constant (table \ref{tab:test_settings}A). To maintain constant animal number density and width of the group, the number of swimmers increased linearly with increasing length of the group. Second, to test the impact of animal number density on the resulting flow, the number of swimmers and the length of the group were kept constant while increasing the cross-sectional area in the spanwise dimensions (table \ref{tab:test_settings}B). This resulted in an animal number density that decreased with increasing width as $1/W^2$, where $W$ is the width of the group. For each calculation with a selected set of parameters, three iterations of swimmers were placed randomly with these specifications while maintaining a minimum nearest neighbour distance of one BL

\begin{table}
  \centering
  \begin{tabular}{lcccc}
    \toprule
     A)&\multicolumn{4}{c}{Length tests}\\
     \midrule
      Test&Number (N)& Length (L) & Width (W) & Number density\\ \hline
        &swimmers& body length& body length& swimmers/body length$^3$\\ \hdashline
     1&40& 4& 5& 0.4\\
     2&100& 10& 5& 0.4\\
     3&200& 20& 5& 0.4\\
     4&400& 40& 5& 0.4\\
     5&520& 52& 5& 0.4\\
      && & &\\
      \bottomrule
     \end{tabular}
     \quad
     \begin{tabular}{lcccc}
     \toprule
     B)&\multicolumn{4}{c}{Animal number density tests}\\
     \midrule
      Test&Number (N)& Length (L) & Width (W) & Number density\\ \hline
        &swimmers& body length& body length& swimmers/body length$^3$\\ \hdashline
    1&100& 20& 2.2&1\\
    2&100& 20& 3&0.6\\
    3&100& 20& 4&0.3\\
    4&100& 20& 7& 0.1\\
    5&100& 20& 10& 0.05\\
    6&100& 20& 19& 0.01\\
    \bottomrule
  \end{tabular}
  \caption{Parameters to be examined are the number of swimmers in the group, $N$, the length of the group, $L$, and the width of the group, $W$. Together, these three parameters result in a group metric that we refer to as the animal number density, measured in animals per BL$^3$ and calculated as $N/(W^2L)$. These parameters are used to examine the impact of changes in A) group length, and B) animal number density.}
  \label{tab:test_settings}
\end{table}

\section{Results}

\subsection{Swimmer response to light and background flow} \label{sec:results_individual}

\begin{figure}
 \centering{\includegraphics[width=\textwidth]{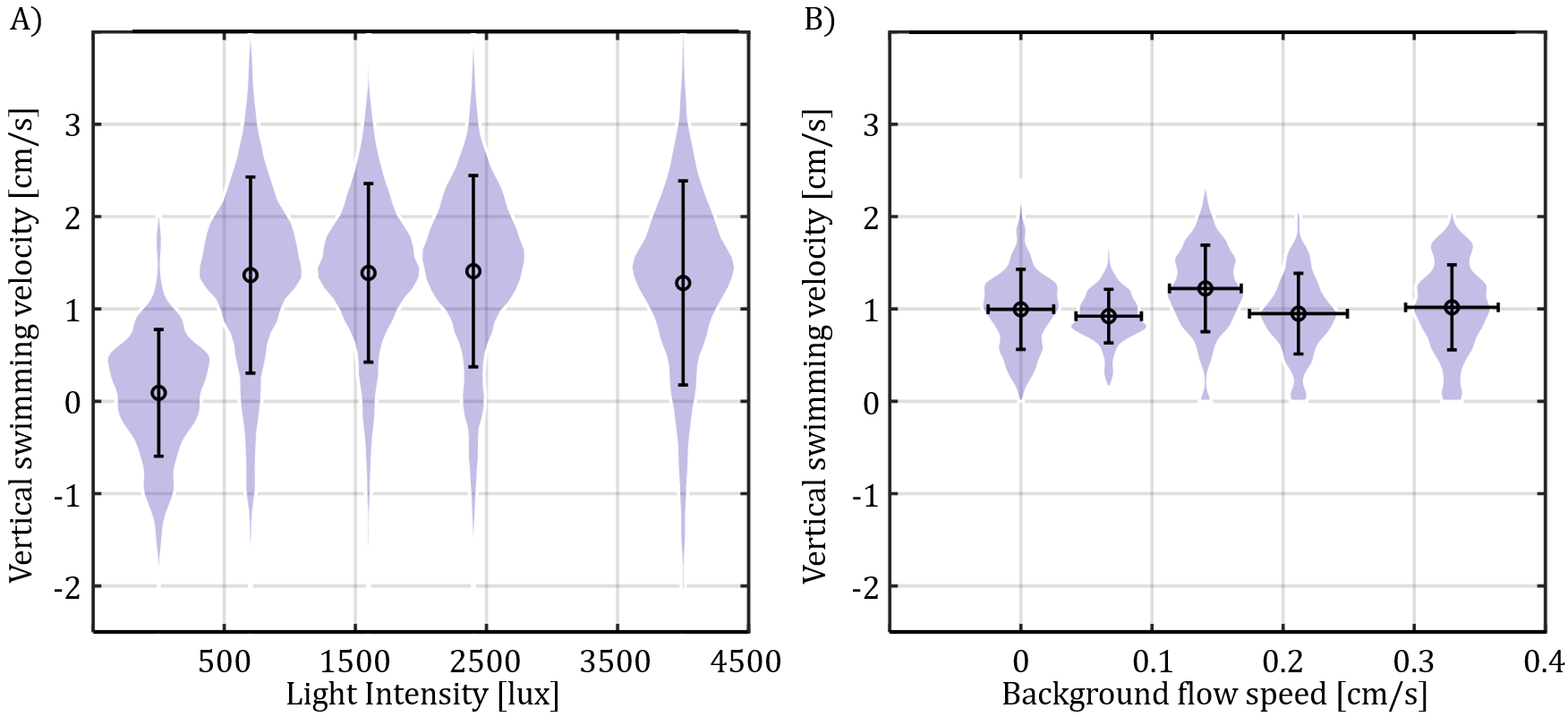}}
 \caption{Brine shrimp maintain a constant vertical velocity under varying environmental conditions. A) Distribution, mean and standard deviation of vertical swimming speed as a function of light intensity, measured in lux. The control condition, with only an infrared lamp (0 lux), is also shown for comparison. B) Distribution, mean and standard deviation of vertical swimming speed as a function of background flow. Horizontal error bars represent range of flows experienced.}
\label{fig:individual}
\end{figure}

Swimmers involved in vertical migration patterns are subject to varying degrees of light exposure and background flow, influenced by the presence of upstream swimmers that obstruct the light source and create wakes. However, brine shrimp consistently maintained swimming speeds irrespective of flow conditions and light intensities tested (figure \ref{fig:individual}). Consequently, in subsequent simulations, swimmers were posited to maintain constant velocity. 

\subsection{Collective swimming dynamics} \label{sec:results_collective}

\begin{figure}
 \centering{\includegraphics[width=\textwidth]{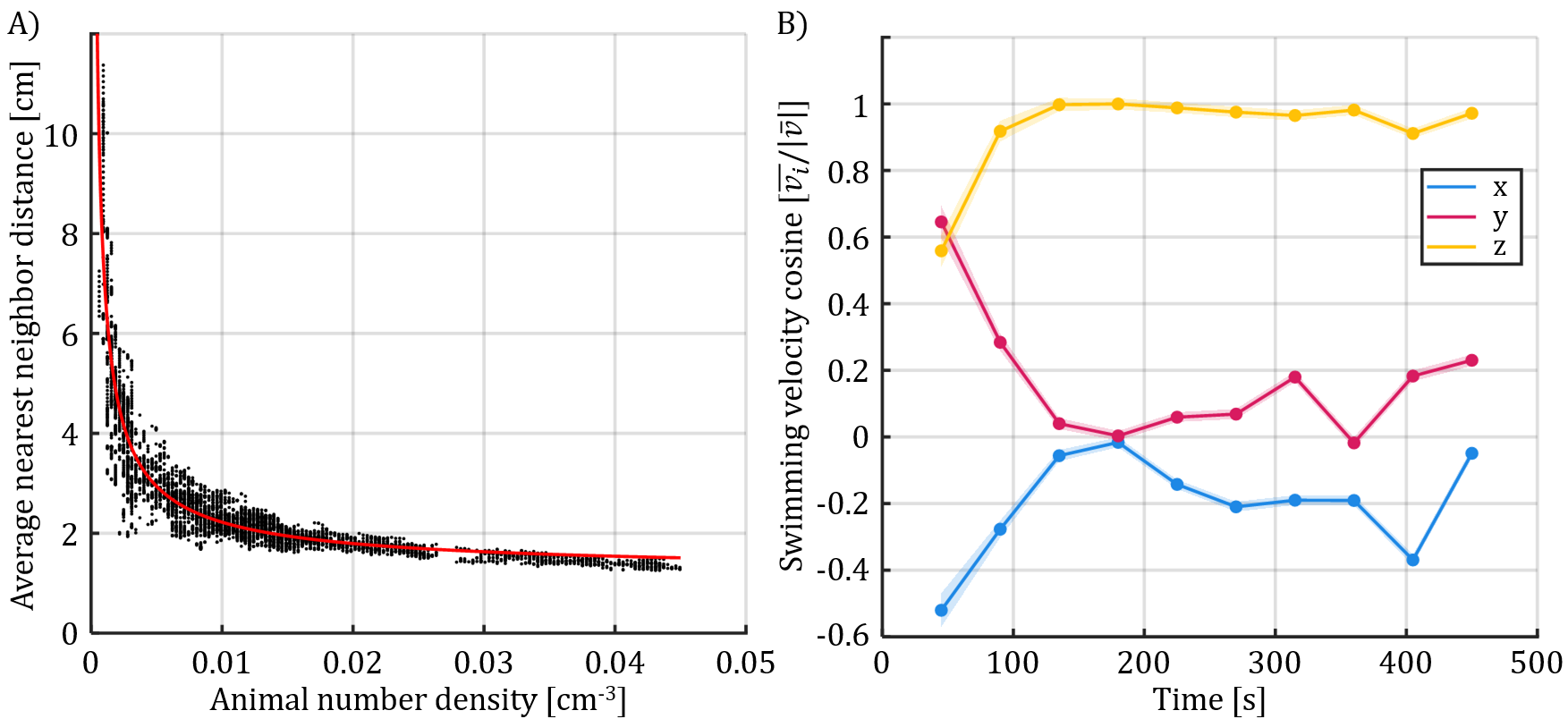}}
 \caption{Data extracted from 3-D brine shrimp trajectories. A) Changes in average nearest neighbour distance during an induced vertical migration with increasing number of brine shrimp within the scanned volume. Power law best fit plotted in red, $y=0.031x^{0.74}+1.16$, with $R^2$ value of 0.85. B) Average swimming velocity components from brine shrimp trajectories over the course of an induced migration, with shaded areas representing the standard error. The target flashlight is located at positive z, above the tank.}
\label{fig:results_3D}
\end{figure}

As the number of swimmers within the scanned volume increases, the average nearest neighbour distance decreases but reaches a limit, evidence of exclusion zones (figure \ref{fig:results_3D}A). To determine the asymptotic limit of the data, we employed the MATLAB fit function to determine the best-fit power law relationship between animal number density, $x$, and the average nearest neighbour distance, $y$). The power law fit was calculated as $y=ax^b+c$. This approach allowed us to investigate how the average nearest neighbour distance converges as the animal number density increases. The asymptotic value, $c$, of the fit was found to be 1.16. Consequently, we approximated the minimum space between swimmers for modelling purposes to be 1 cm.

The components of swimming velocity were computed by first calculating each swimmer's instantaneous velocity based on trajectory data. The velocity was then averaged per individual swimmer over a maximum of 6.5 seconds of data recorded. Next, each swimmer's velocity was averaged across trials at each time step, which were spaced 45 seconds apart. The average velocity vector for each time step is then normalized by the magnitude of the average velocity to arrive at the velocity cosines (\ref{fig:results_3D}B). The dominance of the positive z component indicates a strong, consistent upward motion among the swimmers towards the target flashlight (located at positive z). Thus, we may treat momentum addition as entirely in the z-axis for modelling.

\subsection{Modelled collective hydrodynamics}\label{sec:results_model}

The parameters derived experimentally above were used to inform the wake models for the individual swimmers. These modeled wakes were then applied to various swimmer configurations using the wake superposition framework from \S\ref{sec:superposition} to characterize the collectively induced flow. All calculations were done in the swimmer-fixed reference frame. However, for the sake of clarity, the results presented here are depicted and discussed in the laboratory-fixed frame. 

\subsubsection{Dependence on aggregation size}

\begin{figure}
 \centering{\includegraphics[scale=0.29]{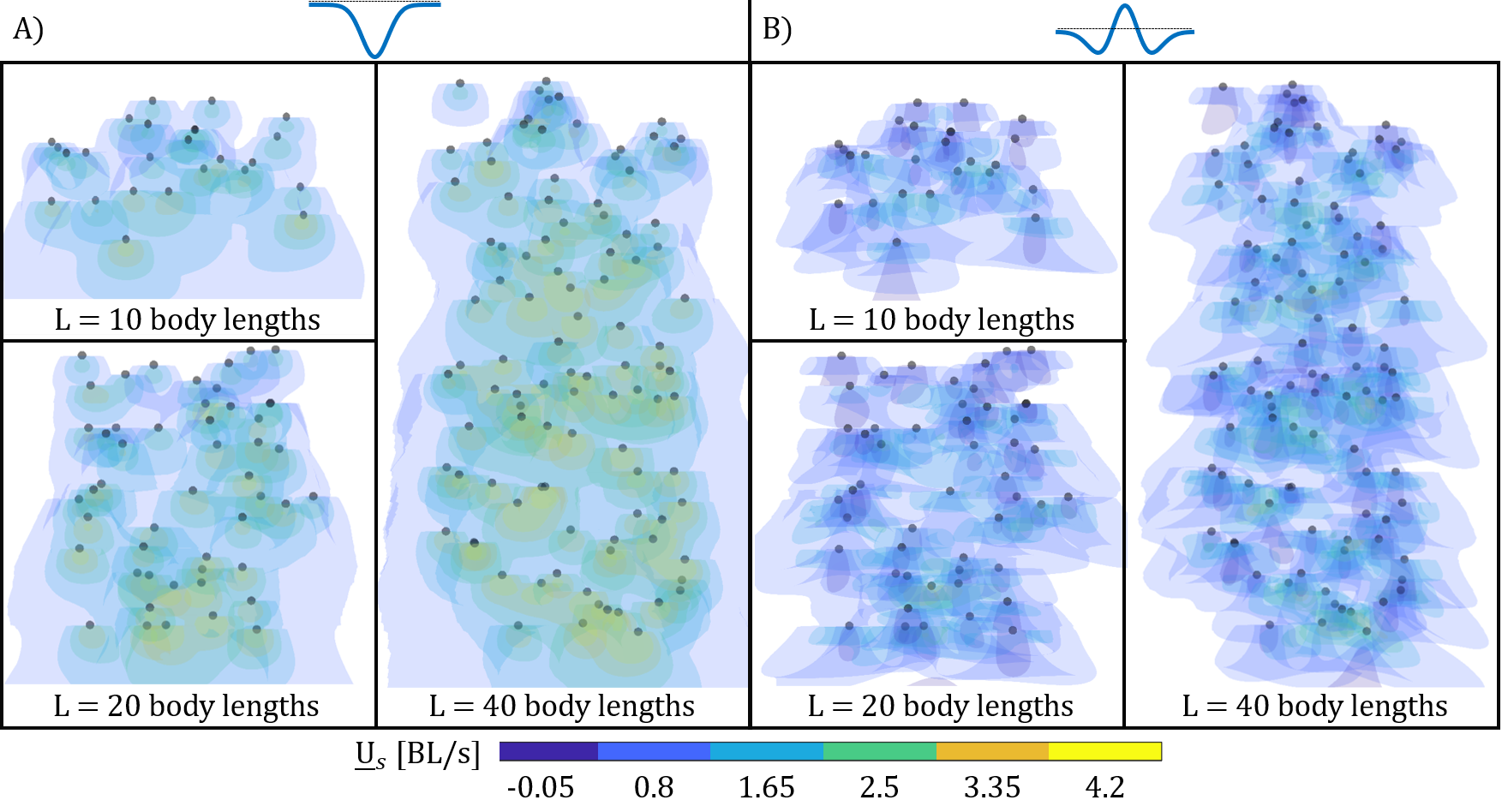}}
 \caption{The 2-D projection of group geometry and induced flow contour map for groups with an animal number density of 0.2 animals per BL$^3$, shown for increasing group lengths. Black spheres represent positions of swimmers. Five isosurfaces of the 3-D flow field output generated from the semianalytical model are superimposed, with the colour indicating the flow magnitude. Results are shown side by side from A) the Gaussian model and B) the wavelet model.}
\label{fig:length}
\end{figure}

\begin{figure}
 \centering{\includegraphics[scale=0.29]{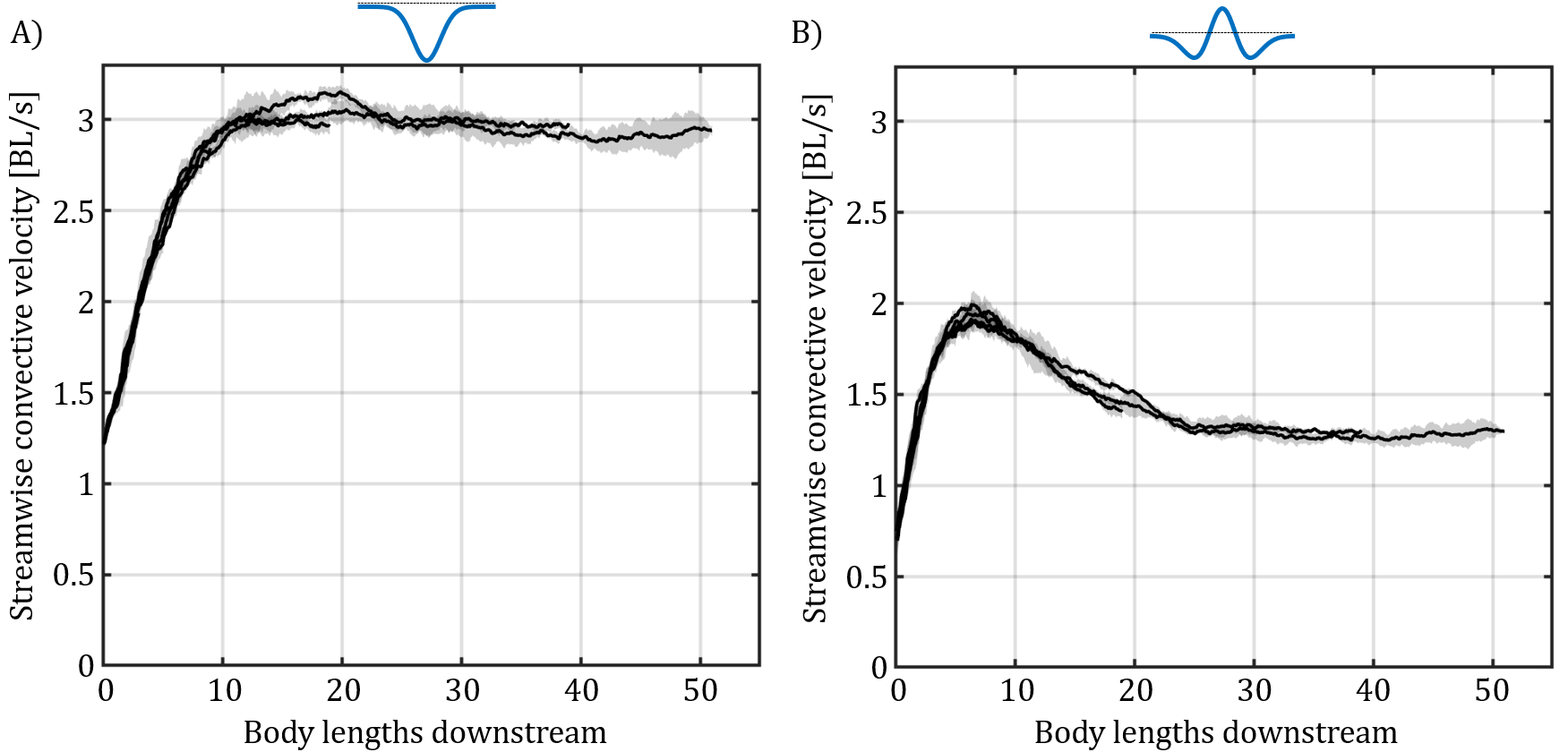}}
\caption{Laboratory frame convection velocity, $U_c(z)-U_\infty$ measured in BL s$^{-1}$, plotted against streamwise distance from start of group, $z$. Each plot represents three randomized iterations, with the line indicating the average value and the shaded areas indicating the standard deviation. Results for five group lengths (4, 10, 20, 40, and 52 BL) are superimposed for comparison using A) the Gaussian model and B) the wavelet model.}
\label{fig:results_length}
\end{figure}

In the first set of calculations, we examine the flow induced by groups with the same animal number density, 0.4 animals per BL$^3$, but different lengths (figure \ref{fig:length}). Three configurations of each group length were generated with randomly placed swimmers. The average convection velocity and standard deviation of each group length was plotted (figure \ref{fig:results_length}). The convection velocity generated within the groups were found to overlap each other. This indicated that the upstream portion of the flow generated within a group was not affected by the downstream flow. Furthermore, the induced flow ceased to exhibit a discernible dependence on the group length beyond a certain threshold, estimated at around 20-30 BL in this case. Consequently, we found that the dynamics of both longer and shorter groups can be approximated by studying the flow generated by any group longer than this threshold length. 

\subsubsection{Dependence on swimmer spacing}

\begin{figure}
 \centering{\includegraphics[scale=0.29]{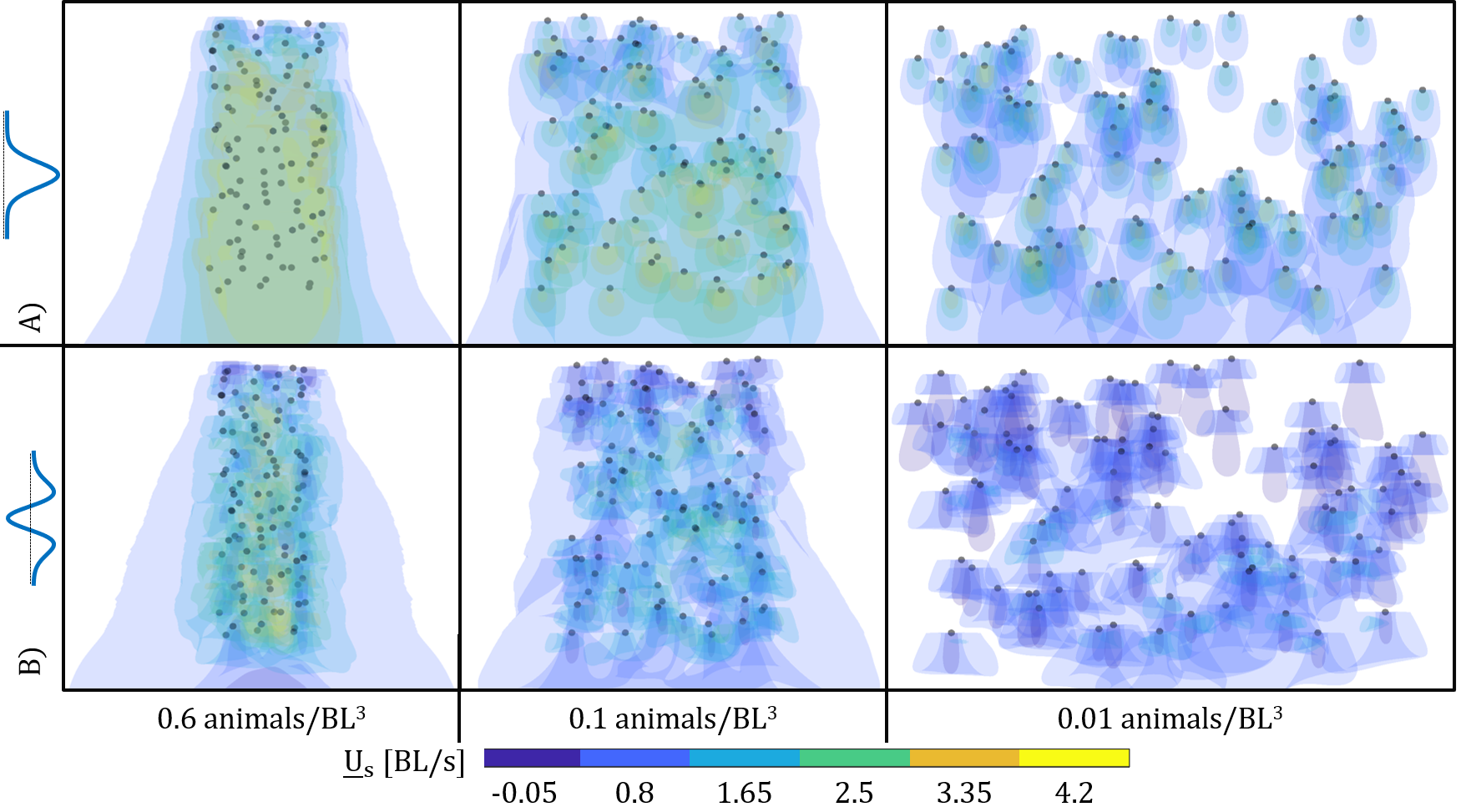}}
 \caption{2-D projection of group geometry and induced flow contour map for 100 swimmers with increasing group width, resulting in decreasing animal number density. Black spheres represent positions of swimmers. Five isosurfaces of the 3-D flow field output generated from the semianalytical model are superimposed, with the colour indicating the flow magnitude. Results are shown side by side from A) the Gaussian model and B) the wavelet model.}
\label{fig:method_APD}
\end{figure}

\begin{figure}
 \centering{\includegraphics[scale=0.29]{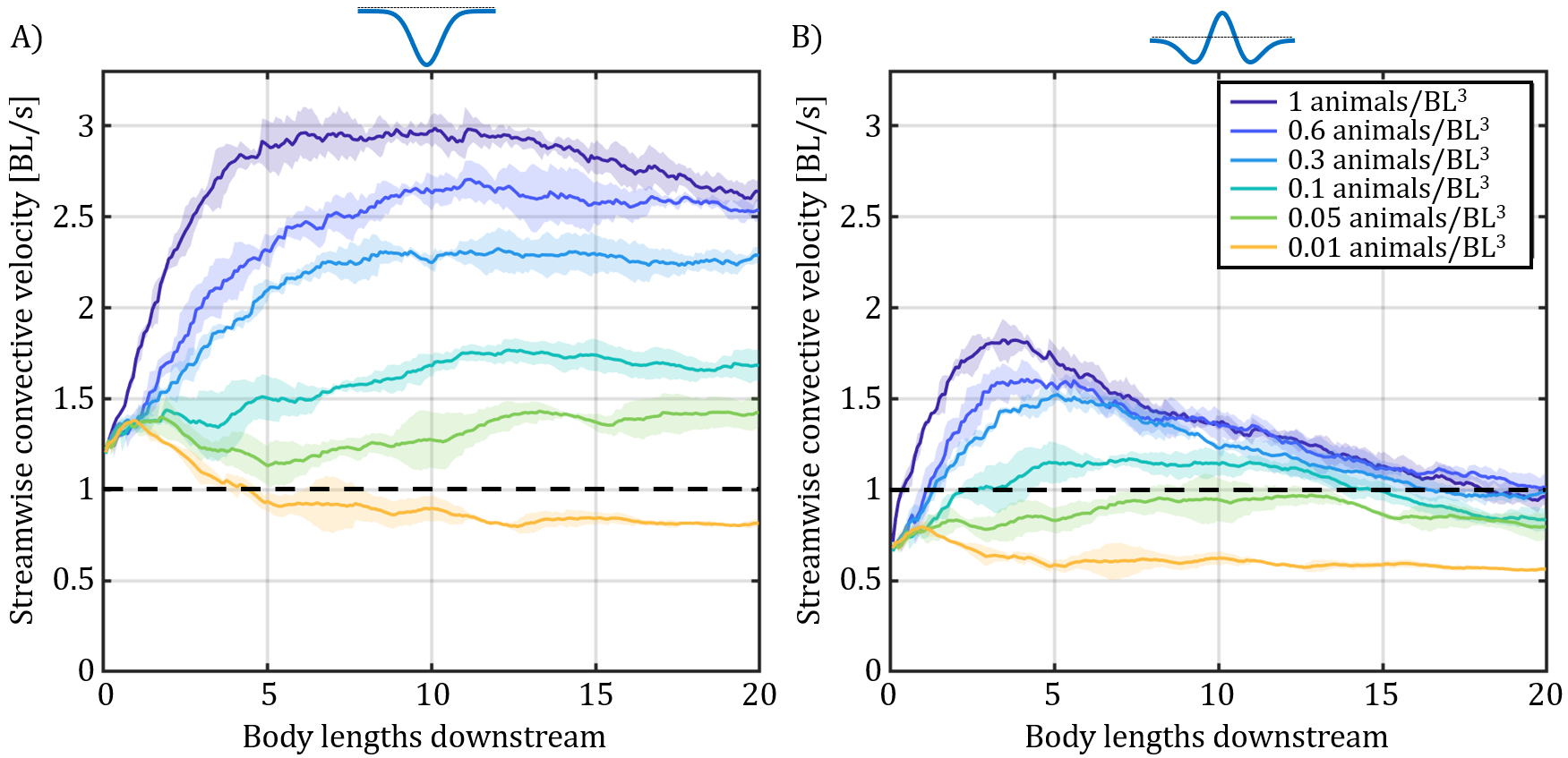}}
 \caption{Laboratory frame convection velocity, $U_c(z)-U_\infty$ measured in BL s$^{-1}$, plotted against streamwise distance from start of group, $z$. Each plot represents three randomized iterations, with the line indicating the average value and the shaded areas indicating the standard deviation. Six animal number densities (0.01, 0.05, 0.1, 0.3, 0.6, and 1 animal per BL$^3$) are plotted with A) the Gaussian model and B) the wavelet model. DA dashed line indicates the swimming speed prescribed in the model, set at 1 BL s$^{-1}$.}
\label{fig:results_density}
\end{figure}

In the second set of simulations, we investigate the influence of swimmer spacing on the flow generated by the collective. We randomly placed 100 swimmers within a volume of constant length but varying widths, resulting in changes in animal number densities (figure \ref{fig:method_APD}). Three simulations were initiated for each case. Although the truncated swarm length results in less distinct asymptote values, a positive correlation between induced convective flow and animal number density was observed (figure \ref{fig:results_density}). Specifically, for groups with animal number densities exceeding 0.05 animals per BL$^3$ using the Gaussian model, a consistent trend emerged: the flow increased steadily with length until reaching a threshold length, beyond which the dependence on length decreased to a near-stable state. Groups with animal number densities below 0.05 animals per BL$^3$ with the Gaussian model and all number densities for the wavelet model exhibited peak flow early in the aggregation process, followed by substantial decreases in flow. For the sparsest cases, 0.01 animals per BL$^3$, the flow at 25 BL downstream was lower than that generated at the beginning of the aggregation. In all cases, at some threshold length, the dependence of flow on length is greatly reduced.

We also observe that in many cases the estimated convection velocity exceeds the velocity prescribed to swimmers in this model, which was set at 1 BL s$^{-1}$. Although this model captures an instantaneous snapshot in time for a specific configuration of swimmers, in reality, swimmers facing a flow exceeding 1 BL s$^{-1}$ would be pushed in the opposite direction to their swimming motion. Thus, these configurations are paradoxical since we have initialized a configuration of swimmers that creates a flow that would make this animal number density impossible to maintain. 

\begin{figure}
 \centering{\includegraphics[scale=0.29]{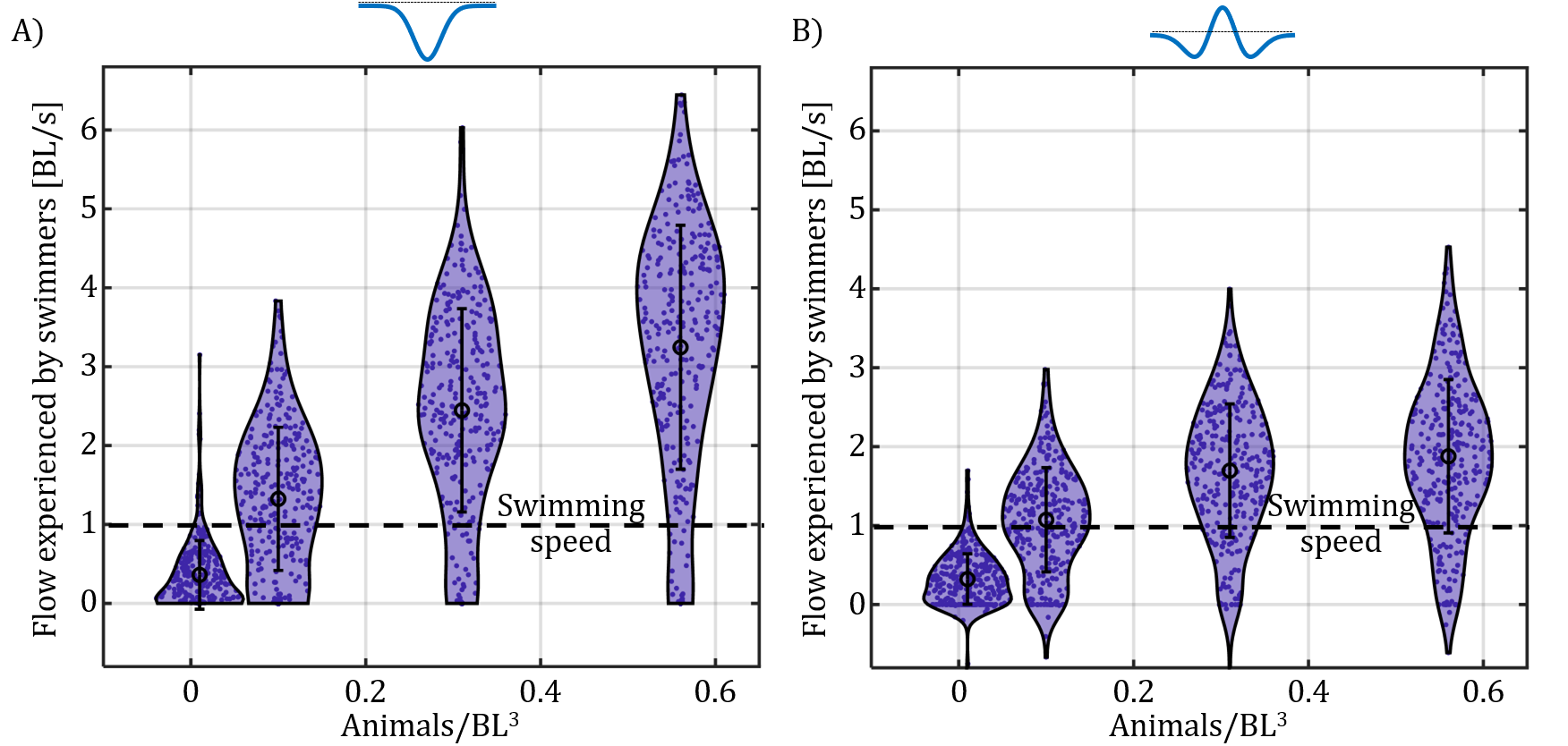}}
 \caption{Distribution of flow experienced by swimmers, $u_0^i-U_\infty$, at different animal number densities. The distributions are shown using A) the Gaussian model and B) the wavelet model.}
\label{fig:results_stability}
\end{figure}

To further investigate the stability of the aggregation, we analysed the flow experienced by individual swimmers within the collective. The distribution of flow velocities for varying animal number densities (figure \ref{fig:results_stability}) revealed that a significant proportion of swimmers in groups denser than 0.1 animals per BL$^3$ experience a flow exceeding 1 BL s$^{-1}$ per second. The magnitude and range of these flows experienced, especially in denser groups, indicate an unsustainable configuration. In this type of individual scale analysis, we observe that in flows generated with the wavelet wake model, some swimmers experience a negative flow, getting a boost by being in the drag-dominated region of an upstream swimmer. It follows that it is also important to examine the flows experienced in relation to the position within the swarm, which may contribute to the range of flow velocities observed between swimmers.

\subsubsection{Comparison with experimental data}

\begin{figure}
 \centering{\includegraphics[scale=0.29]{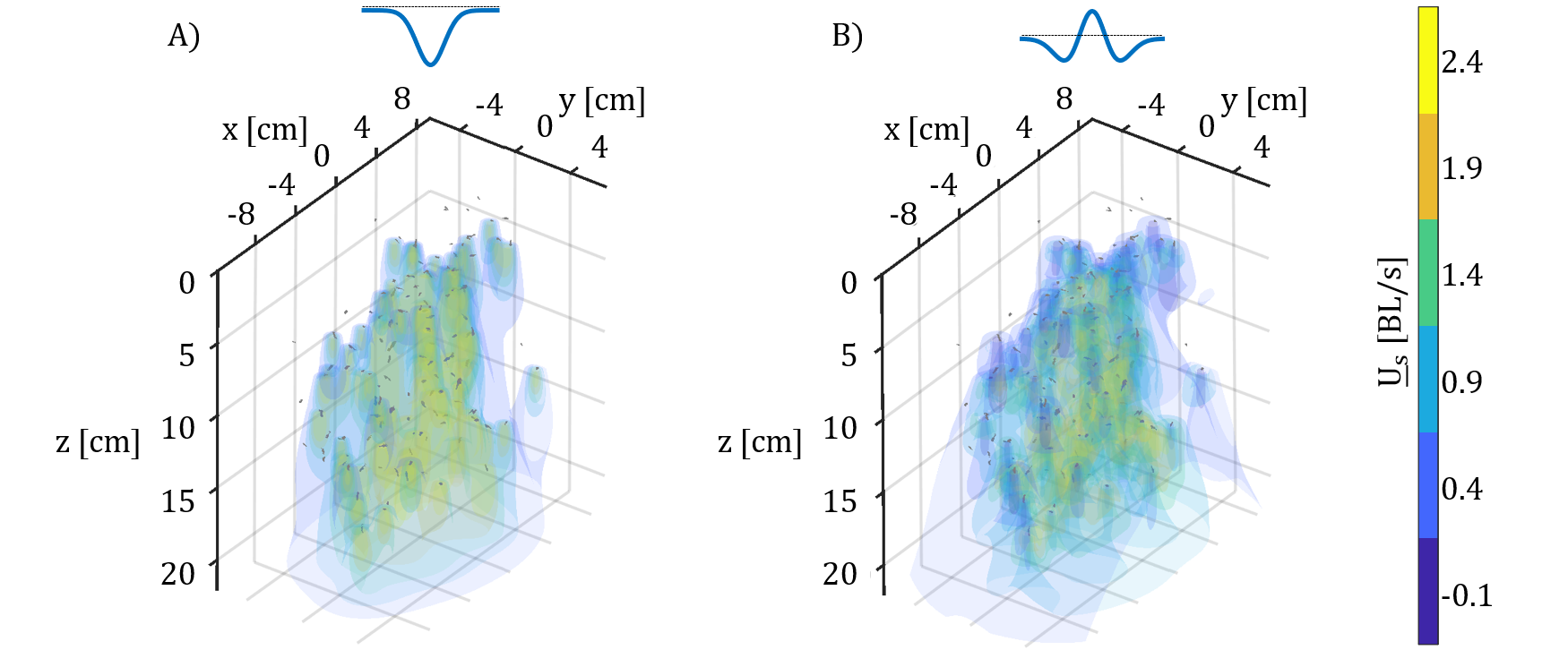}}
 \caption{Scanned and reconstructed brine shrimp from 3-D PTV overlaid with flow field, $U_s(x,y,z)$ generated by A) the Gaussian wake model and B) the wavelet wake model.}
\label{fig:results_exp_overlay}
\end{figure}

\begin{figure}
 \centering{\includegraphics[scale=0.29]{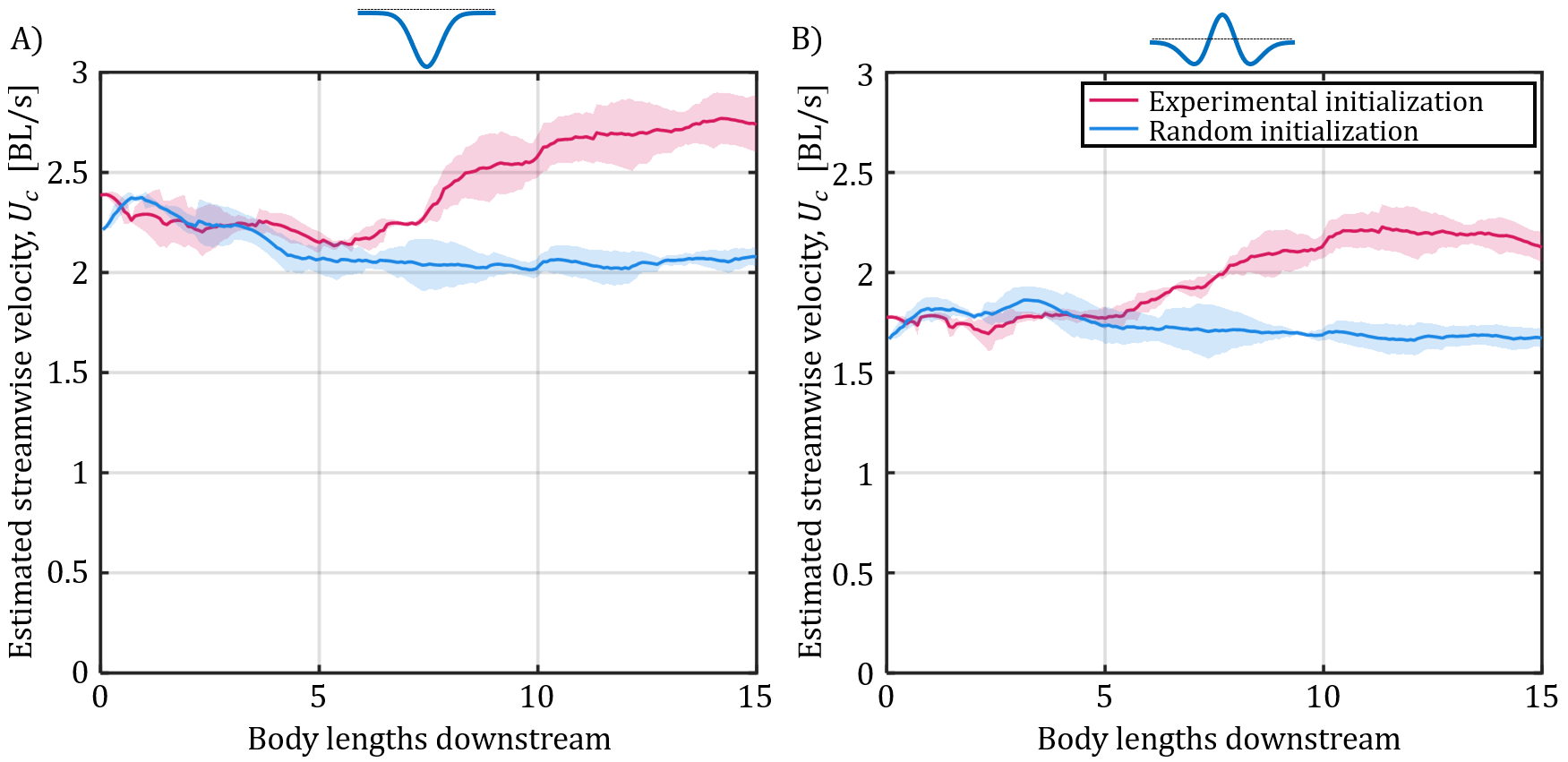}}
 \caption{Comparing convective velocity generated by randomly distributed swimmer locations and experimentally initialized swimmer locations using A) the Gaussian model and B) the wavelet model. Estimated convection velocity plotted against streamwise distance from start of group, $z$ for randomized simulations and for a simulation initialized with locations of brine shrimp during induced vertical migration.}
\label{fig:results_exp_Uc}
\end{figure}

In all previous simulations, the swimmers were randomly placed within specified parameters. To provide context to these findings, we conducted a test by initializing the computational model with three swimmer configurations obtained by 3-D PTV of induced brine shrimp migrations from section \ref{sec:collective}. We used three cases in which 100 brine shrimp were scanned and reconstructed within the volume ($L$= 22 cm, $W_1$= 6.6 cm, $W_2$= 22 cm), resulting in an animal number density of 0.03 animals per cm$^3$ (figure \ref{fig:results_exp_overlay}). This was compared with three simulations initialized with the same volume and number of swimmers, placed randomly. Figure \ref{fig:results_exp_Uc} shows that the flow derived from the simulation with experimentally obtained swimmer configuration resulted in a higher convection magnitude than the randomly initialized simulations. 

\begin{figure}
 \centering{\includegraphics[scale=0.29]{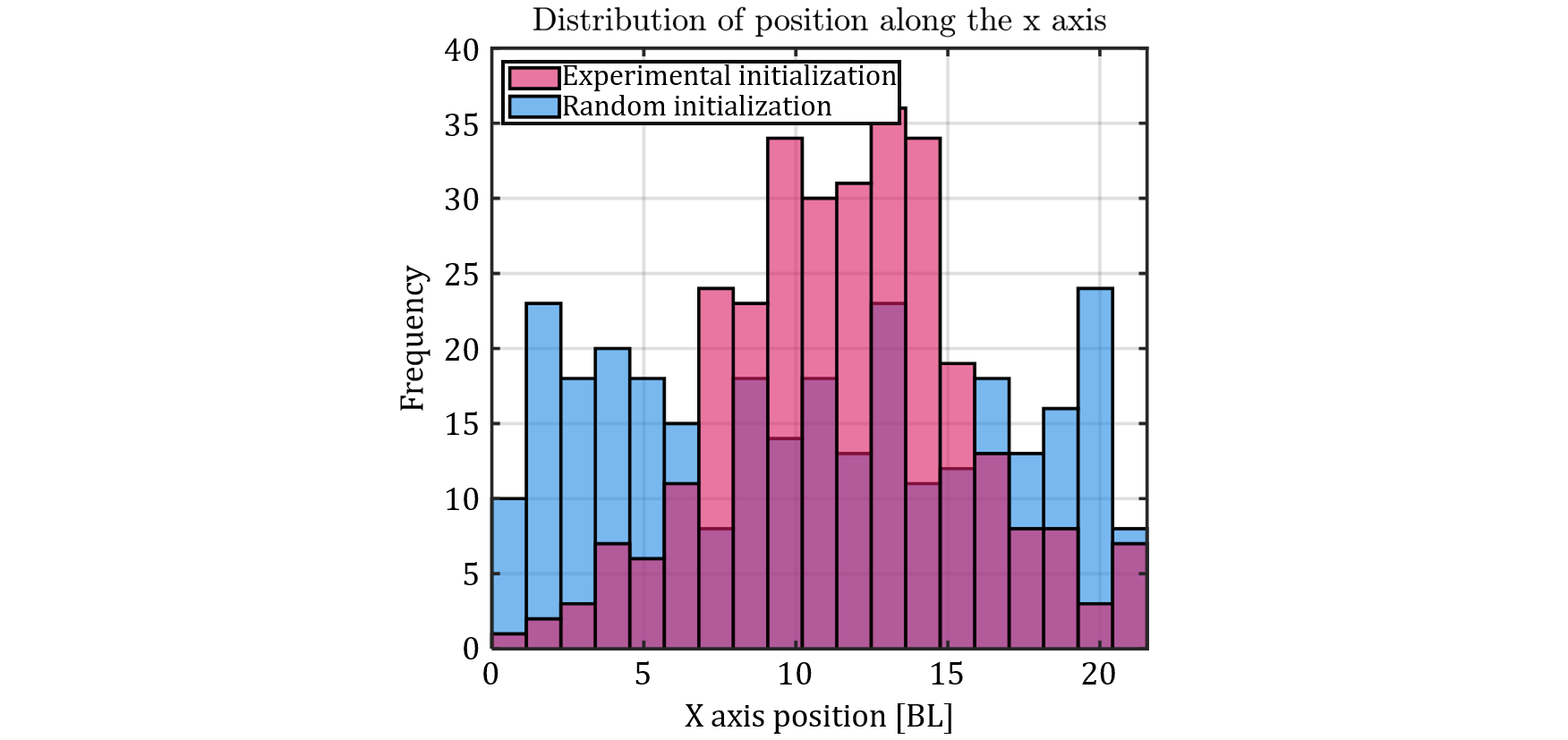}}
 \caption{Comparison of the distribution of swimmer positions between experimentally initialized (pink bars) and randomly initialized groups (blue bars). Swimmers in the experimental set-up are concentrated in the centre, while those initialized randomly are distributed more evenly throughout the volume. The illuminated positions in the tank correspond roughly to the region between $X= [7,13]$.}
\label{fig:results_pos_dist}
\end{figure}

A portion of this difference can be attributed to the fact that the animal number density is overly generalized, failing to capture the spatial variations in the swimmers' configuration. As shown in figure \ref{fig:results_pos_dist}, the concentration of swimmers is significantly higher towards the centre of the tank when experimentally initialized. The increased concentration in the centre of the volume likely results in more frequent swimmer wake interactions, leading to larger induced flow. While it may not be surprising, given that the flashlight was centrally positioned in the tank, attracting the brine shrimp towards the light, it is noteworthy that the swimmers do not avoid or alter their swimming paths to mitigate the higher flow regions created by these interactions.

\begin{figure}
 \centering{\includegraphics[scale=0.29]{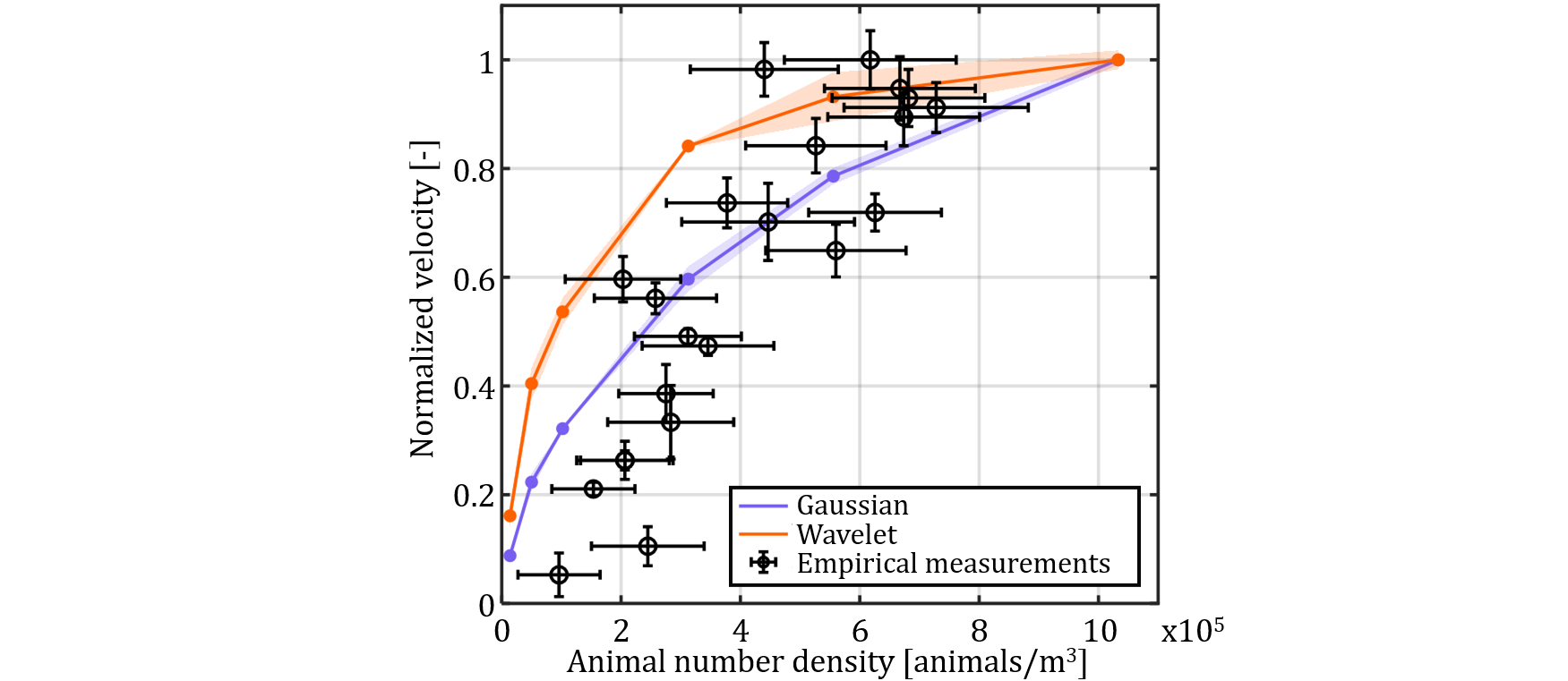}}
 \caption{Comparison of normalized velocity induced by varying densities of swimmers between experimental data and model predictions. Normalized flow velocity as a function of animal number density, comparing experimental data from \citet{Houghton_Dabiri_2019} (black circles with standard deviation bars) with the average and standard deviation (shaded area) from triplicate simulations at corresponding densities using the Gaussian and wavelet models.}
\label{fig:results_comp_exp}
\end{figure}

To further explore the relationship between animal number density and the induced flow velocity, we plotted induced flow from experimental data in \citet{Houghton_Dabiri_2019} against the computational simulations on a normalized scale (figure \ref{fig:results_comp_exp}). This comparison highlights the scaling behaviour of the induced flow as a function of animal number density. While the magnitudes differ, the general trend is consistent across the experimental and simulated data. The power law fit was calculated as $y=ax^b+c$, where $b$ controls the rate at which induced flow velocity changes as a function of animal number density. By comparing the values of $b$ between the experimental ($b=0.61$) and simulated datasets (Gaussian, $b=0.5$; wavelet, $b=0.3$), we see that the Gaussian wake model produces flow magnitude growth rates with animal number density.

\section{Discussion}\label{Discussion}

We have developed an analytical wake superposition model for groups of hydrodynamically interacting organisms. This model was implemented numerically with parameters derived from empirical observations of brine shrimp, incorporating observed responses to light and flow as well as 3-D swimming trajectories. Numerical simulations with this model produce a 3-D flow field and an estimated convection velocity. This semianalytical model provides a quantitative framework for understanding hydrodynamic interactions within swimming aggregations at intermediate Reynolds numbers. 

Our findings highlight the intricate interplay between wake kinematics, swimmer spacing, and overall group size and arrangement in inducing flows within swimming collectives. Notably, the wavelet wake model, when compared with the Gaussian wake model, generates lower-magnitude convective velocities, resulting in swimmers within the group experiencing slower flows. The positive flow regions in the wavelet have an annular shape with the maximum flow value reached over a circle in space. In contrast, the Gaussian wake reaches a maximum value at a single point. Thus, the wavelet model has a more spread-out region of positive flow. In addition, the negative flow region was averaged when looking at the convective velocity. The differences between the flow induced by Gaussian and wavelet wake models exemplify the importance of local flow kinematics and thus motivate continuing work to measure and model individual organism-level flows. Compared to experimental data, the wavelet model predicts flow magnitudes in closer quantitative agreement; however, when normalized by the maximum flow, the Gaussian model more effectively captures the dependence on animal number density.

By comparing groups of different lengths, we found that the flow within the group exhibits decreased sensitivity to the length of the group beyond a threshold. With a uniform distribution of swimmers, there was a constant infusion of momentum to the flow with streamwise distance from the start of the group. However, the velocity of the induced flow increases within the group length only until the mass flux term of the momentum balance dominates, and each individual adds less velocity to the flow than those upstream. The impact of this added velocity decreases further with diffusion before reaching downstream swimmers. In addition, we found that the upstream portion of the flow within the group was not affected by the group downstream. Thus, the dynamics of shorter groups can be extracted from the dynamics of groups longer than a certain threshold.

Simulating different group densities, we found that the collective convective velocity increased with the animal number density. Dense configurations resulted in flows that exceeded the swimming speed of the organism, resulting in unstable structures. This observation raises questions about the apparent stability of swimmer aggregations in field observations, where these high-density configurations are known to persist. The contrast between simulated results and observed natural behaviour prompts inspiration for future model improvements to explore mechanisms used by organisms to navigate and thrive in environments characterized by dynamic collective swimming. For example, there is evidence that animals exploit fluid structures to improve locomotion \citep{Weber_Arampatzis_Novati_Verma_Papadimitriou_Koumoutsakos_2020, Oteiza_Odstrcil_Lauder_Portugues_Engert_2017}. In randomized simulations, positions were initialized by placing swimmers in a prescribed volume swimming directly upward. For vertically swimming negatively buoyant swimmers, momentum excess in the vertical direction is a reasonable generalization. To extend this model beyond constant-speed, unidirectional swimming, the impact of non-aligned trajectories and flow-response behaviours is needed. Similar models for wind turbines have found that wake deflection impacts wake spreading and thus affects aggregate flow characteristics \citep{Shapiro_Gayme_Meneveau_2018}. However, applying these methods to animal behaviour and modelling requires further investigation to determine the suitability. The only constraint in swimming placement was the exclusion zones that maintain a minimum nearest neighbour distance. If the model included some parameters to actively optimize swimmer placement, downstream swimmers might seek the drag region of a wavelet wake or avoid peak flows of a Gaussian wake. Continued work to study the stability of these systems could incorporate discrete time step dynamics to investigate how collective flow-inducing systems evolve.

This model is adaptable to different wake profiles and aggregation configurations, allowing future exploration of flows generated by other organisms with different wake profiles and collective behaviours. Furthermore, this model is applicable across a spectrum of ecological and engineering contexts, including active and passive particle systems such as marine snow and multiphase flows. 

\backsection[Supplementary data]{\label{SupMat} Supplementary material will be available upon publication.}

\backsection[Acknowledgements]{The authors gratefully acknowledge M. F. Howland and S. P. Devey for advice on developing and implementing the numerical simulation. Furthermore, the authors would like to thank Caltech mail services, specifically D. A. Goudeau, for support with the logistics of delivering live animals. Lastly, the authors would like to thank the reviewers from the Journal of Fluid Mechanics for their detailed feedback and helpful suggestions that significantly improved the presentation of this work.}

\backsection[Funding]{This work was supported by funding from the US National Science Foundation through the Alan T. Waterman Award and Graduate Research Fellowship Grant No. DGE 1745301.}

\backsection[Declaration of interests]{The authors report no conflict of interest.}

\backsection[Data availability statement]{All data generated and discussed in this study are available within the article and its supplementary files, or are available from the authors upon request.}

\backsection[Author ORCIDs]{N. Mohebbi, https://orcid.org/0000-0003-4014-6111; J. Hwang, https://orcid.org/0000-0002-9897-7930; M. K. Fu, https://orcid.org/0000-0003-3949-7838; J. O. Dabiri, https://orcid.org/0000-0002-6722-9008}

\appendix
\section{}\label{appA}
We model the spatial evolution of the propulsive jet as a self-similar, axisymmetric jet,
\begin{equation}
  u_s = u_0 \;c(z)\;\xi\left(\frac{r}{\delta(z)}\right). \label{eqn:a_u_s}
\end{equation}

 In the following derivation $u_0$, $\xi(r/\delta(z))$ and $F_z$ are prescribed, and \ref{eqn:a_u_s} is plugged into simplified momentum,

 \begin{equation}
F_z = \rho \iint_{wake} {u_w(x,y,z)} u_s(x,y,z) \;dxdy,\label{eqn:a_basic_momn}
\end{equation}

to arrive at

\begin{equation}
  F_{z} = \rho \iint u_0^2 \;\left(c(z)\;\xi(r,z) - \left[c(z)\;\xi(r,z)\right]^2\right) \;drd\theta, \label{a_F_c_f}
\end{equation}

to derive an expression for $c(z)$ that conserves momentum by definition. Next, $\xi(r/\delta(z))$ is defined as follows for each of the two wake models, and $\delta(z)$ is defined to be the standard deviation at each value of z, $\sigma(z)$. Note that when these functions are used in signal processing and analysis, a normalized version is used, ensuring that the total energy or power is conserved across different scales, which is important for accurate signal analysis. In this derivation, the prefactor $c(z)$ is constructed to conserve momentum directly, negating the need for commonly used normalization prefactors. To solve for $c_g(z)$ and $c_w(z)$, the prefactors for the Gaussian and wavelet wake models, respectively, the two wake shapes,

\begin{equation}
  \xi_{gaussian}\left(\frac{r}{\sigma(z)}\right)= e^{-\frac{r^2}{2\sigma(z)^2}},\label{a_xi_gaus}
\end{equation}

\begin{equation}
  \xi_{wavelet}\left(\frac{r}{\sigma(z)}\right)= -\left(1 - \frac{r^2}{2\sigma(z)^2}\right) \: e^{-\frac{r^2}{3\sigma(z)^2}}\label{a_xi_wave}
\end{equation}

are substituted into \ref{a_F_c_f},

\begin{equation}
  F_{z} = \rho \iint u_0^2 \;\left(c_g(z) e^{-\frac{r^2}{2\sigma(z)^2}} - \left(c_g(z) e^{-\frac{r^2}{2\sigma(z)^2}} \right)^2\right) \;drd\theta, 
\end{equation}

\begin{equation}
  F_{z} = \rho \iint u_0^2 \;\left(c_w -\left(1 - \frac{r^2}{2\sigma(z)^2}\right) \: e^{-\frac{r^2}{3\sigma(z)^2}} - \left(c_w(z) -\left(1 - \frac{r^2}{2\sigma(z)^2}\right) \: e^{-\frac{r^2}{3\sigma(z)^2}}\right)^2\right) \;drd\theta
\end{equation}

 and then integrated in the streamwise direction over a circular cross-section with infinite radius. For each shape, there are two solutions for $c(z)$. We select the solution that is negative and tends to 0 with increasing positive z values,

\begin{equation}
  c_g(z) =1 - \sqrt{1 + \frac{F_{z}}{\pi \rho \sigma(z)^2 u_0^2}} 
  \label{eqn:a_c_x_gaussian},
\end{equation}

\begin{equation}
  c_w(z) = \frac{4}{5}-\frac{4}{5}\sqrt{1 + \frac{5 F_{z}}{12 \pi \rho \sigma(z)^2 u_0^2}}.
  \label{eqn:a_c_x_wavelet}
\end{equation}

Combining \ref{a_xi_gaus} with \ref{eqn:a_c_x_gaussian} and \ref{a_xi_wave} with \ref{eqn:a_c_x_wavelet} and plugging into \ref{eqn:a_u_s} to derive the final expressions for $u_s$, as follows: 

\begin{equation}
  u_{s,wavelet}(x,y,z)= u_0 \left( \frac{4}{5}-\frac{4}{5}\sqrt{1 + \frac{5 F_{z}}{12 \pi \rho \sigma(z)^2 u_0^2}}\right) \left(-1 \left(- \frac{r^2}{2\sigma(z)^2}\right) \: e^{-\frac{r^2}{3\sigma(z)^2}}\right), \label{a_u_s wavelet}
\end{equation}

\begin{equation}
  u_{s,gaussian}(x,y,z)= u_0 \left(1 - \sqrt{1 + \frac{F_{z}}{\pi \rho \sigma(z)^2 u_0^2}} \right) e^{-\frac{r^2}{2\sigma(z)^2}}. \label{a_u_s gaussian}
\end{equation}

Finally, substituting the above equations \ref{a_u_s wavelet} and \ref{a_u_s gaussian} into the wake convection velocity, $u_c$, as follows:

\begin{equation}
{u_c(z)} = \frac{\displaystyle\iint_{wake} {u_w(x,y,z)} u_s(x,y,z) \;dxdy}{\displaystyle\iint_{wake} u_s(x,y,z) \;dxdy}, \label{eqn:a_uc}
\end{equation}

we arrive at

\begin{equation}
u_{c, wavelet}(z) = \frac{u_0}{2} \left(1+ \sqrt{1+ \frac{5 F_z}{12 \pi \rho \sigma(z)^2 u_0^2}}\right), \label{a_u_c wavelet}
\end{equation}

\begin{equation}
u_{c, gaussian}(z) = \frac{u_0}{2} \left(1+ \sqrt{1+ \frac{F_z}{\pi \rho \sigma(z)^2 u_0^2}}\right). \label{a_u_c gaussian}
\end{equation}

\bibliographystyle{jfm}
\bibliography{_manuscript}

\end{document}